\documentclass{cernrep}

\newcommand{\transpose}{^{\textrm{T}}}

\begin{document}
\title{Low-Emittance Storage Rings}
\author{A.~Wolski}
\institute{University of Liverpool, Liverpool, UK}
\maketitle

\begin{abstract}
The effects of synchrotron radiation on particle motion in storage rings are
discussed.  In the absence of radiation, particle motion is symplectic, and the
beam emittances are conserved.  The inclusion of radiation effects in a classical
approximation leads to emittance damping: expressions for the damping times
are derived.  Then, it is shown that quantum radiation effects lead to excitation
of the beam emittances.  General expressions for the equilibrium longitudinal and
horizontal (natural) emittances are derived.  The impact of lattice design on the
natural emittance is discussed, with particular attention to the special cases of
FODO, achromat, and TME style lattices.  Finally, the effects of betatron coupling
and vertical dispersion (generated by magnet alignment and lattice tuning errors)
on the vertical emittance are considered.
\end{abstract}

\section{Introduction}

\begin{figure}[t]
\begin{center}
\includegraphics[width=0.8\textwidth]{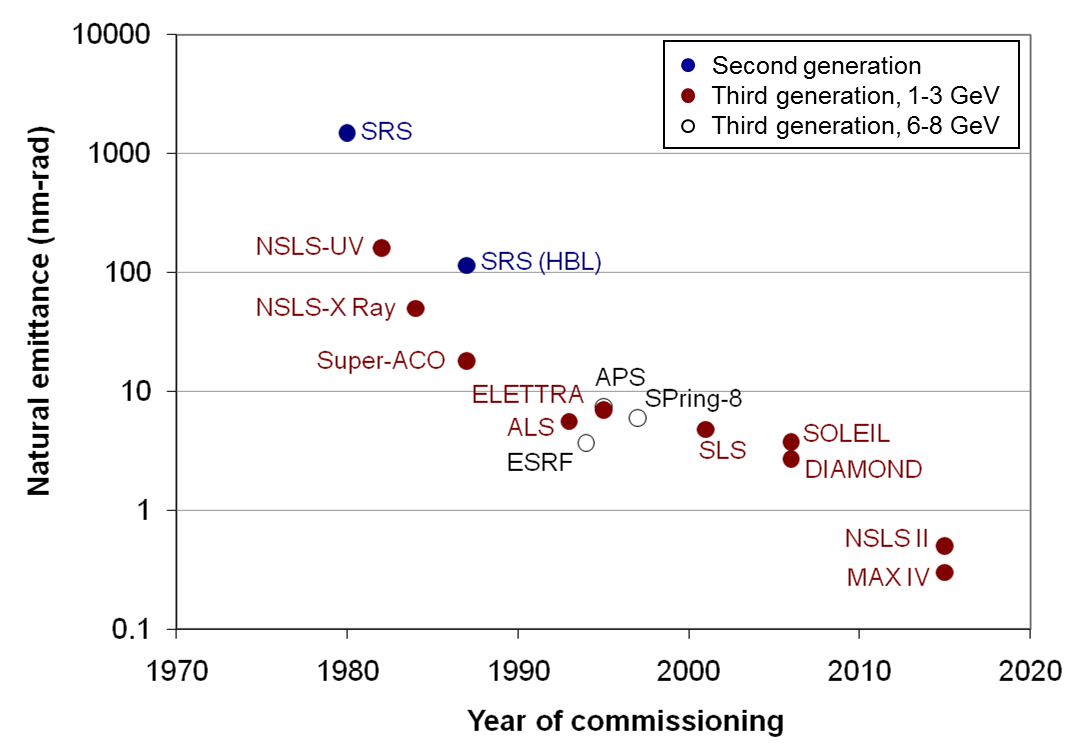}
\caption{Natural emittance of a number of synchrotron light sources.  Reductions
in natural emittance have been driven by the need to produce higher radiation beam
brightness for users.}
\label{figlightsourceemittance}
\end{center}
\end{figure}

Beam emittance in a storage ring is an important parameter for characterising
machine performance.  In the case of a light source, for example, the brightness
of the synchrotron radiation produced by a beam of electrons is directly
dependent on the horizontal and vertical emittances of the beam and is one of
the main figures of merit for users.  Second generation light sources had
natural emittances of order 100\,nm.  Over the years, significant improvements in
lattice designs have been achieved (see Fig.~\ref{figlightsourceemittance}),
motivated largely by user requirements; third generation light sources now
typically aim for natural emittances of just a few nanometres.
In the case of colliders for high energy physics, one of the main figures
of merit is the luminosity, which is a measure of the rate of particle collisions.
Lower emittances allow smaller beam sizes at the interaction point,
leading to higher particle density in the colliding bunches, and higher luminosity
for the same total number of particles in the beam.

There are of course ways of improving the brightness of a light source and the
luminosity of a collider without reducing the emittances: in both cases, for example,
the beam current could be increased.  However, beam currents are generally limited
by collective effects such as impedance-driven instabilities, Touschek scattering,
or (for colliders) beam-beam effects.  Designing and operating a storage ring for
maximum performance involves a good understanding and control of effects
that impact the beam emittances.

In this note, we shall consider the emittance of electron (and positron) storage
rings: because of synchrotron radiation effects, lepton storage rings are able to
achieve very small emittances (of order 1\,nm horizontal emittance, and less than
10\,pm vertical emittance).  We shall begin in Section~\ref{section1} by reviewing
some of the key features of beam dynamics in the absence of synchrotron radiation.
In particular, an important property of the dynamics in such cases is that the particle
motion is symplectic: this has the consequence that the beam emittances (which
characterise the phase space volume occupied by the particles in a beam) are conserved
as the beam moves around the storage ring.  We shall then show that, in a classical
approximation, radiation effects lead to damping of the emittances.  We shall
derive expressions for the exponential damping times.  Then, we shall discuss
how quantum effects of synchrotron radiation lead to excitation of the beam
emittances.  As a result, the emittances of beams in electron (or positron) storage
rings reach equilibrium values determined by the beam energy and lattice design. 

In Section~\ref{section2} we shall apply the expression for the natural
emittance derived in Section~\ref{section1} to particular styles of lattice
design.  In particular, we shall consider FODO, double bend achromat,
theoretical minimum emittance, and multi-bend achromat lattices.  Double bend
achromats are of particular interest for light sources, because they achieve low
natural emittance (leading to high brightness) while providing long, dispersion-free
(or low dispersion) straight sections that are ideal locations for insertion devices
such as undulators or wigglers \cite{clarke2004}.  Insertion devices are useful for
providing intense beams of synchrotron radiation with specific properties.

In a planar storage ring, the vertical emittance is dominated by alignment
and tuning errors, rather than by the design of the lattice.  In Section~\ref{section3}
we shall discuss how the vertical emittance is related to a range of errors,
including steering errors, tilt errors on quadrupoles and vertical alignment
errors on sextupoles.  Betatron coupling and vertical dispersion are important
features of the dynamics in this context, and both will be discussed.
Optimisation of a lattice design for a low-emittance storage ring will
generally involve simulations to characterise the sensitivity of the vertical
emittance to different types of machine error.  For this, techniques are
needed for accurate computation of the equilibrium emittances from models
in which different errors can be included.  We shall consider three techniques
that are widely used for emittance computation, discussing the envelope method
in particular in some detail.  Finally, we shall mention briefly some of the issues
associated with operational tuning of a storage ring for low-emittance operation.


\section{Beam dynamics with synchrotron radiation\label{section1}}

In this section, we shall review the relevant aspects of beam dynamics needed
for understanding the effects of synchrotron radiation.  Our focus will be on
electron (or positron) synchrotron storage rings.  Initially, we shall neglect
radiation effects; then, we shall include the emission of synchrotron radiation
as a perturbation to the motion of individual particles.  This approach is valid if
radiation effects are relatively weak, which means that the energy lost by a particle
through radiation in one synchrotron period should be small compared to the
total energy of a particle.  This is almost invariably the case for practical
storage rings.  We shall consider only incoherent synchrotron radiation,
in other words we shall assume that the motion of each particle and the
radiation that it produces can be considered independently of all other
particles in the beam.  In some regimes (including, for example, in free
electron lasers) particles generate radiation coherently, leading to a strong
enhancement of the radiation produced by a beam.  Generally, some special
efforts are needed to achieve the generation of coherent synchrotron
radiation with sufficient intensity that it has a measurable impact on the
beam; we shall not discuss such situations here.

Briefly, we shall proceed as follows.  The symplectic motion of particles in
an accelerator (i.e.~motion neglecting synchrotron radiation and collective
effects) is conveniently described using action-angle variables.  We shall
define these variables, and use them to review the key features of particle
motion in synchrotron storage rings.  We shall then include the effects of
synchrotron radiation, initially in a classical approximation, leading to expressions
for the energy lost per turn in a storage ring, and the damping times for
the horizontal, vertical and longitudinal emittances.  Finally, we shall
discuss the effects of quantum excitation, and derive results for the
equilibrium beam emittances.  These results will be used in Section~\ref{section2},
where we consider how the equilibrium emittances are affected by
the lattice design in a storage ring.

\subsection{Symplectic motion}

\begin{figure}[b!]
\begin{center}
\includegraphics[width=0.5\textwidth]{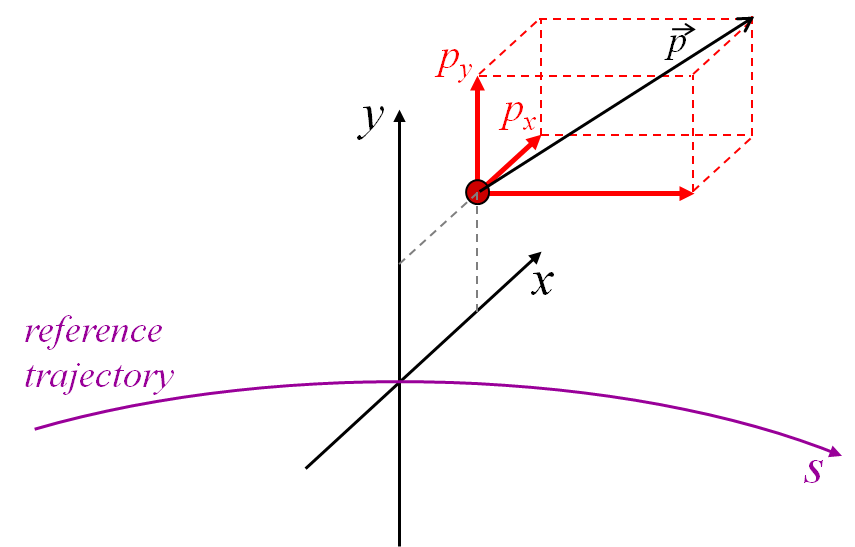}
\caption{Co-ordinate system in an accelerator beam line.  The reference trajectory
can be defined arbitrarly, but is generally chosen so that it describes the trajectory of
a particle with momentum equal to the reference momentum $P_0$.  The distance
along the reference trajectory is parameterised by the independent variable $s$.
At any point along the reference trajectory, the transverse position of a particle
is specified by Cartesian co-ordinates in a plane perpendicular to the reference
trajectory.}
\label{figcoordinatesystem1}
\end{center}
\end{figure}

We work in a co-ordinate system based on a \emph{reference trajectory}
that we define for our own convenience (see Fig.~\ref{figcoordinatesystem1}).
The distance along the reference
trajectory is specified by the \emph{independent variable} $s$.  For simplicity,
in a planar storage ring, the reference trajectory is generally chosen to be
a straight line (passing through the centres of all quadrupole and higher-order
multipole magnets) everywhere except in the dipoles.  In the dipoles, the
reference trajectory follows the arc of a circle with radius $\rho$, such that:
\begin{equation}
B \rho = \frac{P_0}{q},
\end{equation}
where $B$ is the dipole field, $P_0$ is the \emph{reference momentum}
(i.e.~the momentum of particles for which the storage ring is designed)
and $q$ is the particle charge.  $B\rho$ is the \emph{beam rigidity}.

At any point along the reference trajectory, the position of a particle
is specified by the $x$ and $y$ co-ordinates in a plane perpendicular to the
reference trajectory.  We follow the convention in which $x$ is the horizontal
(transverse) co-ordinate, and $y$ is the vertical co-ordinate.

To describe the motion of a particle, we need to give the components of the
momentum of a particle, as well as its co-ordinates.  In the transverse directions
(i.e.~in a plane perpendicular to the reference trajectory) we use the
\emph{canonical momenta} \cite{goldstein2001} scaled by the reference
momentum $P_0$:
\begin{eqnarray}
p_x & = & \frac{1}{P_0} \left(
\gamma m \frac{dx}{dt} + q A_x
\right), \\
p_y & = & \frac{1}{P_0} \left(
\gamma m \frac{dy}{dt} + q A_y
\right).
\end{eqnarray}

Here, $m$ and $q$ are the mass and charge of the particle, $\gamma$ is
the relativistic factor for the particle, and $A_x$ and $A_y$ are the $x$
and $y$ components respectively of the electromagnetic vector potential.
The transverse dynamics are described by giving the transverse co-ordinates and
momenta as functions of $s$ (the distance along the reference trajectory).

\begin{figure}[t]
\begin{center}
\includegraphics[width=0.4\textwidth]{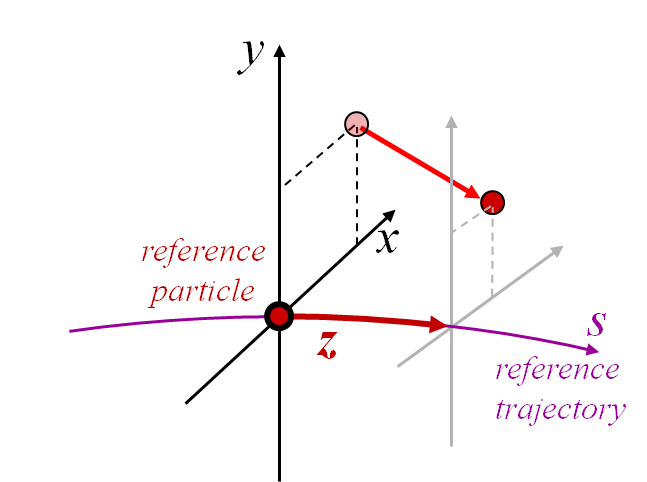}
\caption{Longitudinal co-ordinates in an accelerator beam line.  The
longitudinal co-ordinate $z$ indicates the time that a particle crosses
a plane perpendicular to the reference trajectory at a position $s$
along the reference trajectory.}
\label{figcoordinatesystem2}
\end{center}
\end{figure}

To describe the longitudinal dynamics of a particle, we use a longitudinal
co-ordinate $z$ defined by:
\begin{equation}
z = \beta_0 c(t_0 - t),
\end{equation}
where $\beta_0$ is the normalised velocity of a particle with the reference
momentum $P_0$, $t_0$ is the time at which the reference particle is at
a location $s$, and $t$ is the time at which the particle of interest arrives at
this location.  Physically, the value of $z$ for a particle is approximately
equal to the distance along the reference trajectory between the given particle
and a reference particle travelling along the reference trajectory
with momentum $P_0$ (see Fig.~\ref{figcoordinatesystem2}).
A positive value for $z$ means that the given particle
arrives at a particular location at an earlier time than the reference particle,
i.e.~the given particle is ahead of the reference particle.

The final dynamical variable needed to describe the motion of a particle
is the energy of the particle.  Rather than use the absolute energy or
momentum, we use the \emph{energy deviation} $\delta$, which provides a
measure of the difference between the energy $E$ of a particle and the energy
of a particle with the reference momentum $P_0$:
\begin{equation}
\delta = \frac{E}{P_0 c} - \frac{1}{\beta_0}
= \frac{1}{\beta_0} \left( \frac{\gamma}{\gamma_0} - 1 \right).
\end{equation}
Here, $\gamma_0$ is the relativistic factor for a particle with momentum
equal to the reference momentum.  A particle with momentum equal to the
reference momentum has $\delta = 0$.

Using the above definitions the co-ordinates and momenta form
\emph{canonical conjugate pairs}:
\begin{equation}
(x,p_x), \qquad (y,p_y), \qquad (z,\delta ).
\end{equation}
This means that (continuing to neglect radiation and collective effects) the
equations of motion for particles in an accelerator beam line are given by
Hamilton's equations \cite{goldstein2001}, with an appropriate Hamiltonian
that describes the electromagnetic fields along the beam line.  In a linear
approximation, the change in the values of the variables when a particle
moves along a beam line can be represented by a transfer matrix, $R$:
\begin{equation}
\left( \begin{array}{c}
x \\ p_x \\ y \\ p_y \\ z \\ \delta
\end{array} \right)_{\!\!\!s = s_1}
 = R(s_1 ; s_0) \cdot
\left( \begin{array}{c}
x \\ p_x \\ y \\ p_y \\ z \\ \delta
\end{array} \right)_{\!\!\!s = s_0}
\end{equation}
It is a general property of Hamilton's equations that the transfer matrix $R$
is \emph{symplectic}.  Mathematically, this means that $R$ satisfies the relation:
\begin{equation}
R\transpose  S  R = S,
\label{symplecticcondition}
\end{equation}
where $S$ is the antisymmetric matrix:
\begin{equation}
S =
\left( \begin{array}{cccccc}
 0 & 1 &  0 & 0 &  0 & 0 \\
-1 & 0 &  0 & 0 &  0 & 0 \\
 0 & 0 &  0 & 1 &  0 & 0 \\
 0 & 0 & -1 & 0 &  0 & 0 \\
 0 & 0 &  0 & 0 &  0 & 1 \\
 0 & 0 &  0 & 0 & -1 & 0
\end{array} \right) .
\label{antisymmetricmatrixs}
\end{equation}

The symplectic condition (\ref{symplecticcondition}) imposes strong constraints on the
dynamics.  Physically, symplectic matrices preserve volumes in phase space (this result is
sometimes expressed as Liouville's theorem \cite{goldstein2001}).  For example,
for a linear transformation in one degree of freedom, a particular ellipse in $x$--$p_x$ phase
space will be transformed to an ellipse with (in general) a different shape; but the area of
the ellipse will remain the same.  The number of invariants associated with a linear symplectic
transformation is at least equal to the number of degrees of freedom in the system.  Thus,
for motion in three degrees of freedom, there are at least three invariants.  For particles
in a beam in an accelerator beam line, the invariants are associated with the emittances.
If there is no coupling between the degrees of freedom (so that motion in any
direction $x$, $y$ or $z$ is independent of the motion in the other two directions) then
we can associate an emittance with each of the three co-ordinates, i.e.~there is a
horizontal emittance, a vertical emittance and a longitudinal emittance.  We shall give a
more formal definition of the emittances shortly.

\begin{figure}[t]
\begin{center}
\includegraphics[width=0.5\textwidth]{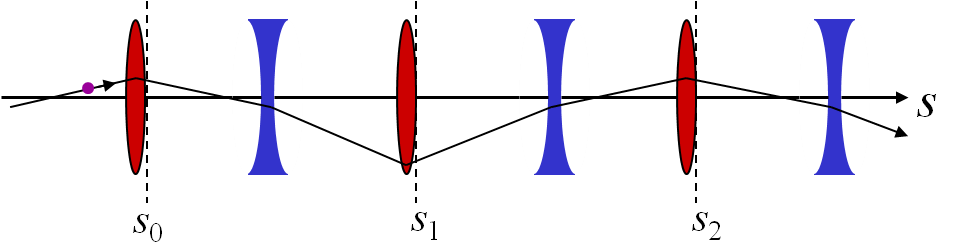} \\
\includegraphics[width=0.4\textwidth]{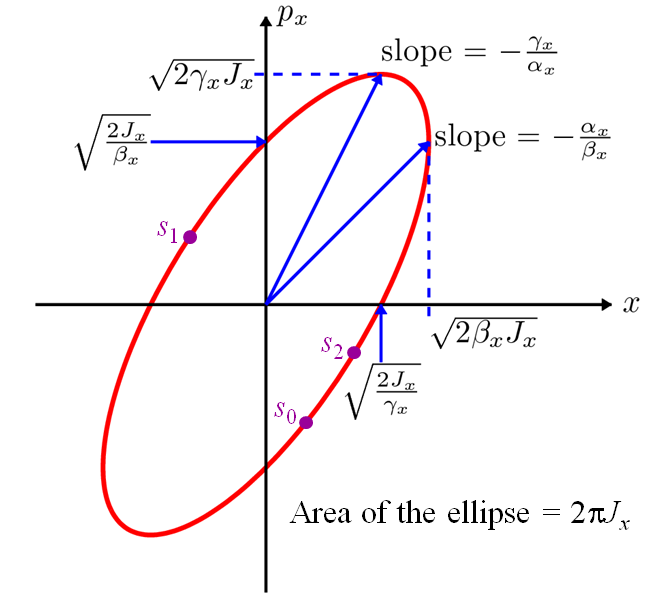}
\caption{Ellipse in phase space defined by plotting the co-ordinate $x$ and
conjugate momentum $p_x$ of a particle after each pass through a unit cell
in a periodic beam line.  The shape of the ellipse is described by the Courant--Snyder
parameters $\alpha_x$, $\beta_x$ and $\gamma_x$; the area of the ellipse is $\pi J_x$,
where $J_x$ is the action variable of the particle.  The shape of the ellipse changes
depending on the chosen starting position within a unit cell; the action remains the same
for any given particle.}
\label{figphasespaceellipse}
\end{center}
\end{figure}

Consider a particle moving through a periodic beam line, without coupling (i.e.~a
beam line with no skew quadrupoles or solenoids).  After each periodic cell, we can plot
the horizontal co-ordinate $x$ and momentum $p_x$ as a point in the horizontal phase
space.  After passing through many cells, observing the particle always at the
corresponding locations in successive cells, and assuming that the motion of the particle
is stable, we find that the points trace out an ellipse
in phase space.  The shape of the ellipse defines the \emph{Courant--Snyder parameters}
\cite{courantsnyder1958}
in the beam line at the observation point: see Fig.~\ref{figphasespaceellipse}.  The area
of the ellipse is a measure of the amplitude of the oscillations. We define the
\emph{horizontal action} $J_x$ of the particle such that the area of the ellipse is equal
to $\pi J_x$.

Applying simple geometry to the phase space ellipse, we find that the action
(for uncoupled motion) is related to the Cartesian variables for the particle by:
\begin{equation}
2J_x = \gamma_x x^2 + 2\alpha_x xp_x + \beta_x p_x^2,
\label{actionvariablex}
\end{equation}
where the Courant--Snyder parameters satisfy the relation:
\begin{equation}
\beta_x \gamma_x - \alpha_x^2 = 1.
\label{betagammaminusalphasquared}
\end{equation}
We define the \emph{horizontal angle} variable $\phi_x$ as follows:
\begin{equation}
\tan \phi_x = -\beta_x \frac{p_x}{x} - \alpha_x.
\label{anglevariablex}
\end{equation}
For a particle with a particular action (i.e.~on an ellipse with a given area) the
angle variable specifies the position of the particle around the ellipse.  The action-angle
variables \cite{goldstein2001} provide an alternative to the Cartesian variables for
describing the dynamics.  Although we have not shown that this is the case, the action-angle
variables form a canonical conjugate pair: that is, the equations of motion expressed
in terms of the action-angle variables can be derived from Hamilton's equations, using
an appropriate Hamiltonian (determined as before by the electromagnetic fields along the
beam line).  The advantage of using action-angle variables to describe particle motion
in an accelerator is that, under symplectic transport (i.e.~neglecting radiation and
collective effects), the action of a particle is constant.  We can of course define
vertical and (in a synchrotron storage ring) longitudinal action-angle variables in
the same way as we defined the horizontal action-angle variables.

The expressions for the action (\ref{actionvariablex}) and the angle (\ref{anglevariablex})
can be inverted, to give expressions for the Cartesian co-ordinate and momentum in
terms of $J_x$ and $\phi_x$:
\begin{eqnarray}
x & = & \sqrt{2\beta_x J_x} \cos \phi_x,
\label{xintermsactionangle} \\
p_x & = & -\sqrt{\frac{2J_x}{\beta_x}} \left( \sin \phi_x + \alpha_x \cos \phi_x \right).
\label{pxintermsactionangle}
\end{eqnarray}

The emittance $\varepsilon_x$ of a bunch of particles can be defined as the average
action of all particles in the bunch:
\begin{equation}
\varepsilon_x = \langle J_x \rangle .
\label{emittancemeanaction}
\end{equation}
For \emph{uncoupled} motion, and assuming that the angle variables of different
particles are uncorrelated, it follows from (\ref{xintermsactionangle}) and
(\ref{pxintermsactionangle}) that the second order moments of the particle distribution
are related to the Courant--Snyder parameters and the emittance:
\begin{eqnarray}
\langle x^2 \rangle & = & \beta_x \varepsilon_x, \\
\langle x p_x \rangle & = & -\alpha_x \varepsilon_x, \\
\langle p_x^2 \rangle & = & \gamma_x \varepsilon_x.
\end{eqnarray}
Using (\ref{betagammaminusalphasquared}), we then find that the emittance can
be expressed in terms of the second order moments as:
\begin{equation}
\varepsilon_x^2 = \langle x^2 \rangle\langle p_x^2 \rangle - \langle x p_x \rangle^2.
\label{emittancerms}
\end{equation}
However, we stress that this relation holds only for uncoupled motion.  The expression
for the emittance (\ref{emittancemeanaction}) can be generalised without too much
difficulty to coupled motion (see, for example \cite{wolski2006}), leading to
\emph{normal mode emittances} that are conserved
under symplectic transport even where coupling is present.  However, the expression for
the emittance (\ref{emittancerms}) is less easily generalised to include coupling, and
an emittance that is defined by (\ref{emittancerms}) will, in general, not be constant
in a beam line where there is coupling.

\subsection{Vertical damping by synchrotron radiation}

So far, we have considered only symplectic transport, i.e.~motion of a
particle in drift spaces or in the electromagnetic fields of dipoles, quadrupoles, rf
cavities etc.~without any radiation.
However, we know that a charged particle moving through an electromagnetic
field will (in general) undergo acceleration, and a charged particle
undergoing acceleration will radiate energy in the form of electromagnetic waves.
We now address the question of the impact that this radiation will have on the
motion of a particle in a synchrotron storage ring.
We shall consider first the case of uncoupled
vertical motion: for a particle in a storage ring, this turns out to be the
simplest case.  Since we are primarily interested in the dynamics of the particles
generating the radiation, we quote a number of results regarding the properties
of the radiation itself (rather than derive these results from first principles).


The first result that we quote for the properties of synchrotron radiation, is that
radiation from a relativistic charged particle is emitted within a cone of opening angle
of $1/\gamma$, where $\gamma$ is the relativistic factor for the particle \cite{jackson1998}.
The axis of the cone is tangent to the trajectory of the particle at the point
where the radiation is emitted.  For an ultra-relativistic particle, $\gamma \gg 1$, and
we can assume that the radiation is emitted directly along the instantaneous direction
of motion of the particle.


Consider a particle with initial momentum $P \approx P_0$, that emits radiation
carrying momentum $dP$.  The momentum of the particle after emitting radiation is:
\begin{equation}
P^\prime = P - dP \approx P \left( 1 - \frac{dP}{P_0} \right).
\end{equation}
Since there is no change in direction of the particle, the vertical component of the
momentum must scale in the same way as the total momentum of the particle:
\begin{equation}
p^\prime_y \approx p_y \left( 1 - \frac{dP}{P_0} \right).
\end{equation}
Now we substitute this into the expression for the vertical betatron action
(valid for \emph{uncoupled} motion):
\begin{equation}
2J_y = \gamma_y y^2 + 2\alpha_y y p_y + \beta_y p_y^2,
\end{equation}
to find the change in the action resulting from the emission of radiation:
\begin{equation}
dJ_y = -\left( \alpha_y y p_y + \beta_y p_y^2 \right) \frac{dP}{P_0}.
\label{djy1}
\end{equation}
Note that in (\ref{djy1}) we neglect a term that is second order in $dP/P_0$.
This term vanishes in the classical approximation when we consider the
emission of an infinitesimal amount of radiation in an infinitesimal time interval $dt$;
however, we shall see later that including quantum effects, the second order term
will lead to excitation of the action.
Retaining for the present only the first order term in $dP/P_0$, averaging (\ref{djy1})
over all particles in the beam gives:
\begin{equation}
\langle dJ_y \rangle = d\varepsilon_y = -\varepsilon_y \frac{dP}{P_0},
\label{depsilonybydt}
\end{equation}
where we have used:
\begin{eqnarray}
\langle yp_y \rangle & = & -\alpha_y \varepsilon_y, \\
\langle p_y^2 \rangle & = & \gamma_y \varepsilon_y,
\end{eqnarray}
and:
\begin{equation}
\beta_y \gamma_y - \alpha_y^2 = 1.
\end{equation}

The emittance is conserved under symplectic transport, so if the
effects of radiation are `slow' (i.e.~the rate of change of energy
from radiation is small compared to the total energy of a particle divided
by the revolution period), then for a particle in a storage ring we can
average the momentum loss around the ring.
From (\ref{depsilonybydt}):
\begin{equation}
\frac{d\varepsilon_y}{dt} = -\frac{\varepsilon_y}{T_0}
\oint \frac{dP}{P_0} \approx
- \frac{U_0}{E_0 T_0} \varepsilon_y
= - \frac{2}{\tau_y} \varepsilon_y,
\end{equation}
where $T_0$ is the revolution period, and $U_0$ is the energy lost through
synchrotron radiation in one turn.
The approximation is valid for an ultra-relativistic particle, which has
$E_0 \approx P_0 c$.
The damping time $\tau_y$ is defined by:
\begin{equation}
\tau_y = 2 \frac{E_0}{U_0} T_0.
\end{equation}
The evolution of the emittance is given by:
\begin{equation}
\varepsilon_y(t) = \varepsilon_y(t = 0) \exp \! \left( -2 \frac{t}{\tau_y} \right).
\label{verticalemittanceevolution}
\end{equation}
Typically, in an electron storage ring, the damping time is of order several
tens of milliseconds, while the revolution period is of the order of a microsecond.
In such a case, radiation effects are indeed slow compared to the revolution
frequency.

Note that we made the assumption that the momentum of the particle was
close to the reference momentum, i.e.~$P \approx P_0$.
If the particle continues to radiate without any restoration of energy, we will
reach a point where this assumption is no longer valid.
However, electron storage rings contain rf cavities to restore the energy lost through
synchrotron radiation.  For a thorough analysis of synchrotron radiation effects on
the vertical motion (at least, with a classical model for the radiation), we should
consider the change in momentum of a particle as it moves through an rf cavity.
However, in general, rf cavities are designed to provide a longitudinal electric field.
This means that particles experience a change in longitudinal momentum
as they pass through a cavity, without any change in transverse momentum.
In other words, the vertical momentum $p_y$ of a particle will remain constant
as the particle moves through an rf cavity, which will therefore have no effect
on the emittance of the beam.


To complete our calculation of the vertical damping time, we need to find
the energy lost by a particle through synchrotron radiation on each turn through
the storage ring.  At this point, we quote a second result from the theory of synchrotron
radiation: the radiation power from a relativistic particle following a circular trajectory of
radius $\rho$ is given by Li\'enard's formula \cite{jackson1998}:
\begin{equation}
P_\gamma = \frac{q^2 c}{6\pi \epsilon_0} \frac{\beta^4 \gamma^4}{\rho^2}
= \frac{C_\gamma c}{2\pi} \frac{c^4 P^4}{\rho^2}
= \frac{C_\gamma}{2\pi} c^5 q^2 B^2 P^2
\approx \frac{C_\gamma}{2\pi} c^3 q^2 B^2 E^2,
\label{synchrotronradiationpower}
\end{equation}
where the particle has charge $q$, velocity $\beta c \approx c$, energy $E = \gamma mc^2$
and momentum $P = \beta\gamma mc$.  The particle travels on a path with radius $\rho$
in a magnetic field of strength $B$.  The approximation in the final expression of
(\ref{synchrotronradiationpower}) is valid for ultra-relativistic particles, $\gamma \gg 1$.
$\epsilon_0$ is the permittivity of free space, and
$C_\gamma$ is a physical constant given by:
\begin{equation}
C_\gamma = \frac{q^2}{3\epsilon_0 (mc^2)^4}.
\end{equation}
For electrons, $C_\gamma \approx 8.846 \times 10^{-5}\,\textrm{m/GeV}^3$.
Note that the radiation power has a very strong scaling with the particle mass: the
larger the mass of the particle, the smaller the amount of radiation emitted.  In proton
storage rings, except at extremely high energy, synchrotron radiation effects are
generally negligible.
For a particle with the reference energy, travelling close to the speed of
light along the reference trajectory, we can find the energy loss by integrating
the radiation power around the ring:
\begin{equation}
U_0 = \oint P_\gamma \, dt \approx \oint P_\gamma \, \frac{ds}{c}.
\end{equation}
Using the expression (\ref{synchrotronradiationpower}) for $P_\gamma$, we find:
\begin{equation}
U_0 \approx \frac{C_\gamma}{2\pi} E_0^4 \oint \frac{1}{\rho^2} \,ds,
\end{equation}
where $\rho$ is the radius of curvature of the particle trajectory, and we
assume that the particle energy is equal to the reference energy $E_0$.
For convenience, we assume that the closed orbit is the same as the reference
trajectory for a particle with the reference momentum.

\begin{figure}[t]
\begin{center}
\includegraphics[width=0.6\textwidth]{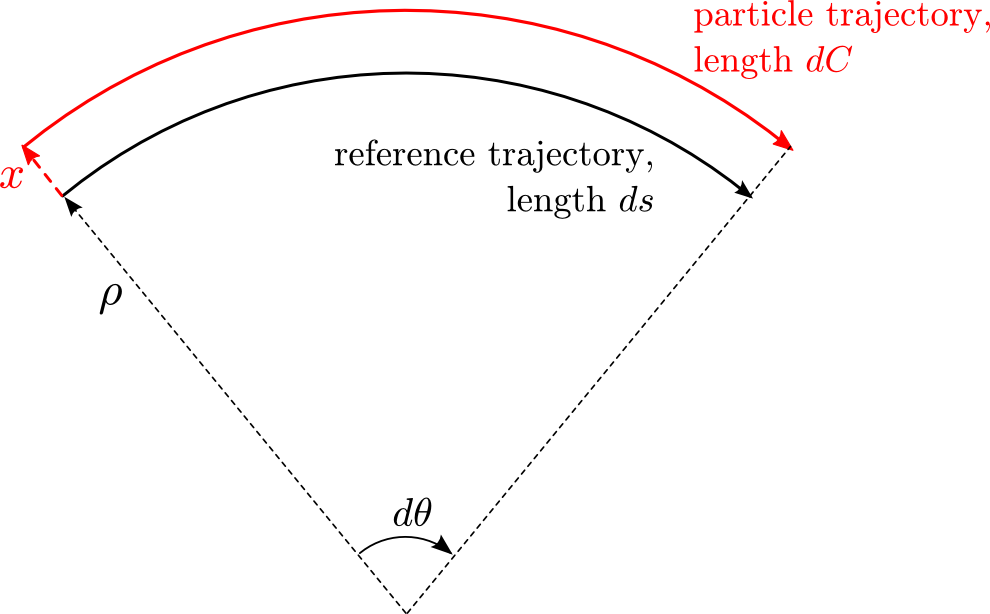}
\caption{Change in path length of a particle following a trajectory
offset from the reference trajectory.  If a particle has co-ordinate $x$
and follows a path parallel to the reference trajectory, then the length
of the path followed by the particle is $dC = (1 + x/\rho) ds$, where
$\rho$ is the radius of curvature of the reference trajectory, and $ds$
is the corresponding distance along the reference trajectory.}
\label{figdispersivepathlength}
\end{center}
\end{figure}

Following convention, we define the \emph{second synchrotron radiation integral},
$I_2$ \cite{sands1970}:
\begin{equation}
I_2 = \oint \frac{1}{\rho^2} \, ds.
\label{eqn_i2}
\end{equation}
In the ultra-relativistic limit,
the energy loss per turn $U_0$ is written in terms of $I_2$ as:
\begin{equation}
U_0 = \frac{C_\gamma}{2\pi} E_0^4 I_2.
\label{energylossperturn}
\end{equation}
Note that $I_2$ is a property of the lattice (actually, a property of the reference
trajectory), and does not depend on the properties of the beam.
Conventionally, there are five synchrotron radiation integrals used to express
the effects of synchrotron radiation on the dynamics of ultra-relativistic particles in an accelerator.
The first synchrotron radiation integral is not, however, directly related
to the radiation effects.
It is defined as:
\begin{equation}
I_1 = \oint \frac{\eta_x}{\rho} \, ds,
\end{equation}
where $\eta_x$ is the horizontal dispersion.
$I_1$ is related to the \emph{momentum compaction factor} $\alpha_p$,
which plays an important role in the longitudinal dynamics, and
describes the change in the length of the closed orbit with respect to particle energy:
\begin{equation}
\frac{\Delta C}{C_0} = \alpha_p \delta + O(\delta^2).
\end{equation}
The length of the closed orbit changes with energy because of
dispersion in regions where the reference trajectory has some curvature
(see Fig.~\ref{figdispersivepathlength}):
\begin{equation}
dC = (\rho + x)\, d\theta = \left( 1 + \frac{x}{\rho} \right) \, ds.
\end{equation}
If $x = \eta_x \delta$, then:
\begin{equation}
dC =\left( 1 + \frac{\eta_x \delta}{\rho} \right) \, ds.
\end{equation}
The momentum compaction factor can be written:
\begin{equation}
\alpha_p = \left. \frac{1}{C_0} \frac{dC}{d\delta} \right|_{\delta = 0}
= \frac{1}{C_0} \oint \frac{\eta_x}{\rho}\, ds = \frac{I_1}{C_0}.
\end{equation}

\subsection{Horizontal damping}

Analysis of the effect of synchrotron radiation on the vertical emittance
was relatively straightforward.  When we consider the horizontal emittance,
there are three complications that we need to address.
First, the horizontal motion of a particle is often strongly coupled to the
longitudinal motion.  We cannot treat the horizontal motion without also
considering (to some extent) the longitudinal motion. Second, where the
reference trajectory is curved (usually, in dipoles), the length of the path
taken by a particle depends on the horizontal co-ordinate with respect to
the reference trajectory.  This can be a significant effect since dipoles
inevitably generate dispersion (a variation of the orbit with respect to
changes in particle energy), so the length of the path taken by a particle
through a dipole will depend on its energy.  Finally, dipole magnets are
sometimes built with a gradient, in which case the vertical field seen by a
particle in a dipole will depend on the horizontal co-ordinate of the particle.


Coupling between transverse and longitudinal planes in a beam line is usually
represented by the dispersion, $\eta_x$ and $\eta_{px}$, defined by:
\begin{eqnarray}
\eta_x & = & \left. \frac{dx_\mathrm{co}}{d\delta} \right|_{\delta = 0}, \\
\eta_{px} & = & \left. \frac{dp_{x,\mathrm{co}}}{d\delta} \right|_{\delta = 0},
\end{eqnarray}
where $x_\mathrm{co}$ and $p_{x,\mathrm{co}}$ are the co-ordinate and
momentum for a particle with energy deviation $\delta$ on a closed orbit.
We use the horizontal action-angle
variables $J_x$ and $\phi_x$ to describe the horizontal betatron oscillations of
a particle with respect to the dispersive closed orbit, i.e.~the closed orbit for
a particle with energy deviation $\delta$.
In terms of the horizontal dispersion and betatron action, the horizontal
co-ordinate and momentum of a particle are given by:
\begin{eqnarray}
x & = & \sqrt{2\beta_x J_x} \cos \phi_x + \eta_x \delta, \\
p_x & = & -\sqrt{\frac{2J_x}{\beta_x}} \left( \sin \phi_x + \alpha_x \cos \phi_x \right)
  + \eta_{px} \delta.
\end{eqnarray}
When a particle emits radiation, we have to take into account both the change
in momentum of the particle, and the change in co-ordinate and momentum
with respect to the new (dispersive) closed orbit.
Note that when we analysed the vertical motion, we assumed that there
was no vertical dispersion.  This is the case in an ideal, planar storage ring,
but as we shall discuss later, alignment errors on the magnets can lead to the
generation of some vertical dispersion that depends on the errors, the
effects of which cannot always be neglected.

Taking all the above effects into account for the horizontal motion, we can
proceed along the same lines as for the analysis of the vertical emittance.
That is, we first write down the changes in co-ordinate $x$ and momentum $p_x$
resulting from an emission of radiation with momentum $dP$ (taking into
account the additional effects of dispersion).
Then, we substitute expressions for the new co-ordinate and momentum into
the expression for the horizontal betatron action, to find the change in action
resulting from the radiation emission.  Averaging over all particles in the
beam gives the change in the emittance that results from radiation emission
from each particle in the beam.
Finally, we integrate around the ring (taking account of changes in path length and
field strength with the horizontal position in the bends) to find the change in
emittance over one turn.


Filling in the steps in this calculation, we proceed as follows.
First, we note that, in the presence of dispersion, the action $J_x$ is
written:
\begin{equation}
2J_x = \gamma_x \tilde{x}^2 + 2 \alpha_x \tilde{x} \tilde{p}_x + \beta_x \tilde{p}_x^2,
\end{equation}
where $\tilde{x}$ and $\tilde{p}_x$ are the horizontal co-ordinate and momentum
with respect to the dispersive closed orbit:
\begin{eqnarray}
\tilde{x} & = & x - \eta_x \delta, \\
\tilde{p}_x & = & p_x - \eta_{px} \delta.
\end{eqnarray}
After emission of radiation carrying momentum $dP$, the variables change by:
\begin{eqnarray}
\delta & \mapsto & \delta - \frac{dP}{P_0}, \\
\tilde{x} & \mapsto & \tilde{x} + \eta_x \frac{dP}{P_0},\\
\tilde{p}_x & \mapsto & \tilde{p}_x \left( 1 - \frac{dP}{P_0} \right) + \eta_{px} \frac{dP}{P_0}.
\end{eqnarray}
We write the resulting change in the action as:
\begin{equation}
J_x \mapsto J_x + dJ_x.
\end{equation}
The change in the horizontal action is:
\begin{equation}
dJ_x = - w_1 \frac{dP}{P_0} + w_2 \left( \frac{dP}{P_0} \right)^{\!2},
\label{photonchangehorizontalaction}
\end{equation}
where, in the limit $\delta \to 0$:
\begin{equation}
w_1 = \alpha_x x p_x + \beta_x p_x^2 - \eta_x (\gamma_x x  + \alpha_x p_x)
 - \eta_{px} (\alpha_x x + \beta_x p_x),
\label{w1}
\end{equation}
and:
\begin{equation}
w_2 = \frac{1}{2} \left( \gamma_x \eta_x^2 + 2\alpha_x \eta_x \eta_{px}
 + \beta_x \eta_{px}^2 \right) - \left( \alpha_x \eta_x + \beta_x \eta_{px} \right) p_x
 + \frac{1}{2} \beta_x p_x^2.
\label{w2}
\end{equation}
Treating radiation as a classical phenomenon, we can take the limit
$dP \to 0$ in the limit of small time interval, $dt \to 0$.
In this approximation, the term that is second order in $dP$ vanishes, and
we can write for the rate of change of the action:
\begin{equation}
\frac{dJ_x}{dt} = -w_1 \frac{1}{P_0} \frac{dP}{dt}
 \approx -w_1 \frac{P_\gamma}{P_0 c},
\label{rateofchangeofaction}
\end{equation}
where $P_\gamma$ is the \emph{rate of energy loss} of the particle through
synchrotron radiation (\ref{synchrotronradiationpower}).
To find the \emph{average} rate of change of horizontal action, we integrate
over one revolution period:
\begin{equation}
\frac{dJ_x}{dt} = -\frac{1}{T_0} \oint w_1 \frac{P_\gamma}{P_0 c} \, dt.
\end{equation}
It is more convenient, given a particular lattice design, to integrate over the
circumference of the ring, rather than over one revolution period.
However, we have to be careful changing the variable of integration (from time $t$
to distance $s$) where the reference trajectory is curved:
\begin{equation}
dt = \frac{dC}{c} = \left( 1 + \frac{x}{\rho} \right) \frac{ds}{c}.
\end{equation}
So:
\begin{equation}
\frac{dJ_x}{dt} = -\frac{1}{T_0 P_0 c^2} \oint w_1 P_\gamma 
\left( 1 + \frac{x}{\rho} \right) \, ds,
\label{dJxdt}
\end{equation}
where the rate of energy loss $P_\gamma$ is given by (\ref{synchrotronradiationpower}).

We have to take into account the fact that in general, the field strength in a dipole can
vary with position.  To first order in $x$ we can write:
\begin{equation}
B = B_0 + x \frac{\partial B}{\partial x}.
\label{gradientdipole}
\end{equation}
Substituting (\ref{gradientdipole}) into (\ref{synchrotronradiationpower}), and with the
use of (\ref{w1}), we find (after some algebra) that, averaging over all
particles in the beam:
\begin{equation}
\oint \left\langle w_1 P_\gamma \left( 1 + \frac{x}{\rho} \right) \right\rangle \, ds
= c U_0 \left( 1 - \frac{I_4}{I_2} \right) \varepsilon_x,
\label{a7}
\end{equation}
where the energy loss per turn $U_0$ is given by (\ref{energylossperturn}), the second synchrotron
radiation integral $I_2$ is given by (\ref{eqn_i2}), and the fourth synchrotron radiation integral
is $I_4$:
\begin{equation}
I_4 = \oint \frac{\eta_x}{\rho} \left( \frac{1}{\rho^2} + 2k_1 \right) ds.
\label{eqn_i4}
\end{equation}
$k_1$ is the normalised quadrupole gradient in the dipole field:
\begin{equation}
k_1 = \frac{q}{P_0} \frac{\partial B_y}{\partial x}.
\end{equation}
Note that in (\ref{eqn_i4}), the dispersion and quadrupole
gradient contribute to the integral only in the dipoles: in other parts of the
ring, where the beam follows a straight path, the curvature $1/\rho$
is zero.

Averaging (\ref{dJxdt}) over all particles in the beam and combining with (\ref{a7}) we have:
\begin{equation}
\frac{d\varepsilon_x}{dt} = - \frac{1}{T_0} \frac{U_0}{E_0}
\left( 1 - \frac{I_4}{I_2} \right) \varepsilon_x.
\end{equation}
Defining the horizontal damping time $\tau_x$:
\begin{equation}
\tau_x = \frac{2}{j_x} \frac{E_0}{U_0} T_0,
\label{horizontaldampingtime1}
\end{equation}
where:
\begin{equation}
j_x = 1 - \frac{I_4}{I_2},
\label{horizontaldampingpartition}
\end{equation}
the evolution of the horizontal emittance can be written:
\begin{equation}
\frac{d\varepsilon_x}{dt} = - \frac{2}{\tau_x} \varepsilon_x.
\label{horizontalemittanceevolution1}
\end{equation}
The quantity $j_x$ is called the \emph{horizontal damping partition number}.
For most synchrotron storage ring lattices, if there is no gradient in the dipoles
then $j_x$ is very close to 1.
From (\ref{horizontalemittanceevolution1}) the horizontal emittance decays exponentially:
\begin{equation}
\varepsilon_x(t) = \varepsilon_x(t=0) \exp \!\left(-2 \frac{t}{\tau_x} \right).
\label{horizontalemittanceevolution}
\end{equation}

\subsection{Longitudinal damping}

So far we have considered only the effects of synchrotron radiation on the
transverse motion, but there are also effects on the longitudinal motion.
Generally, synchrotron oscillations are treated differently from betatron
oscillations because in one revolution of a typical storage ring, particles complete many
betatron oscillations but only a fraction of a synchrotron oscillation.  In other words,
the betatron tunes are $\nu_\beta \gg 1$, but the synchrotron tune is $\nu_s \ll 1$.
To find the effects of radiation on synchrotron motion, we proceed as follows.
We first write down the equations of motion (for the dynamical variables $z$ and $\delta$)
for a particle performing synchrotron motion, including the radiation energy loss.
Then, we express the energy loss per turn as a function of the energy deviation
of the particle.  This introduces a damping term into the equations of motion.
Finally, solving the equations of motion gives synchrotron oscillations (as expected)
with amplitude that decays exponentially.

The changes in energy deviation $\delta$ and longitudinal co-ordinate $z$ for
a particle in one turn around a storage ring are given by:
\begin{eqnarray}
\Delta \delta & = & \frac{qV_\textrm{rf}}{E_0}
\sin \! \left( \phi_s - \frac{\omega_\textrm{rf}z}{c} \right)
- \frac{U}{E_0}, \\
\Delta z & = & - \alpha_p C_0 \delta,
\end{eqnarray}
where $V_\textrm{rf}$ is the rf voltage, $\omega_\textrm{rf}$ the rf frequency,
$E_0$ is the reference energy of the beam, $\phi_s$ is the nominal rf phase,
and $U$ (which may be different from $U_0$) is the energy lost by the particle
through synchrotron radiation.  Strictly speaking, since the longitudinal co-ordinate
$z$ is a measure of the \emph{time} at which a particle arrives at a particular location
in the ring, changes in $z$ with respect to energy should be written in terms of the
\emph{phase slip factor} $\eta_p$, which describes the change in revolution period
with respect to changes in energy, rather than in terms of the momentum compaction
factor $\alpha_p$.  The phase slip factor and the momentum compaction factor are
related by (see, for example \cite{wiedemann2007}):
\begin{equation}
\eta_p = \alpha_p - \frac{1}{\gamma_0^2},
\end{equation}
where $\gamma_0$ is the relativistic factor for a particle with the reference momentum.
But for a storage ring operating a long way above transition (which is the situation we
shall assume here) $\alpha_p \gg 1/\gamma_0^2$, so $\eta_p \approx \alpha_p$.
It is slightly more convenient to work with the momentum compaction factor, since
this depends (essentially) on just the geometry of the lattice and the optical functions
(in particular, the dispersion); whereas the phase slip factor depends also on the
beam energy.

If the revolution period in the storage ring is $T_0$, then we can write the longitudinal
equations of motion for the particle:
\begin{eqnarray}
\frac{d \delta}{dt} & = & \frac{qV_\textrm{rf}}{E_0 T_0}
\sin \! \left( \phi_s - \frac{\omega_\textrm{rf}z}{c} \right)
- \frac{U}{E_0 T_0}, \label{longitudinaleom1} \\
\frac{dz}{dt} & = & - \alpha_p c \delta. \label{longitudinaleom2}
\end{eqnarray}
To solve these equations, we have to make some assumptions.
First, we assume that $z$ is small compared to the rf wavelength:
\begin{equation}
\frac{\omega_\textrm{rf}| z |}{c} \ll 1.
\end{equation}
The synchrotron radiation power produced by a particle depends on the energy
of the particle.
We assume that the energy deviation is small, $| \delta | \ll 1$, so we can work
to first order in $\delta$:
\begin{equation}
U = U_0 + \Delta E \left. \frac{dU}{dE} \right|_{E = E_0}
  = U_0 + E_0 \delta \left. \frac{dU}{dE} \right|_{E = E_0}.
\end{equation}
Finally, we assume that the rf phase $\phi_s$ is set so that for
$z = \delta = 0$, the rf cavity restores exactly the amount of energy lost
by synchrotron radiation.
With these assumptions, the equations of motion become:
\begin{eqnarray}
\frac{d \delta}{dt} & = & - \frac{qV_\textrm{rf}}{E_0 T_0}
\cos ( \phi_s ) \frac{\omega_\textrm{rf}}{c}z
- \frac{1}{T_0} \delta \left. \frac{dU}{dE} \right|_{E = E_0}, \label{zeom1} \\
\frac{dz}{dt} & = & - \alpha_p c \delta. \label{zeom2}
\end{eqnarray}
Taking the derivative of (\ref{zeom1}) with respect to $t$, and substituting for
$dz/dt$ from (\ref{zeom2}) gives:
\begin{equation}
\frac{d^2 \delta}{dt^2} + 2\alpha_E \frac{d\delta}{dt} + \omega^2_s \delta  = 0.
\end{equation}
This is the equation for a damped harmonic oscillator, with frequency $\omega_s$
and damping constant $\alpha_E$ given by:
\begin{eqnarray}
\omega_s^2 & = & -\frac{qV_\textrm{rf}}{E_0} \cos (\phi_s ) \frac{\omega_\textrm{rf}}{T_0} \alpha_p, \\
 & & \nonumber \\
\alpha_E & = & \frac{1}{2T_0} \left. \frac{dU}{dE} \right|_{E = E_0}.
\end{eqnarray}
If $\alpha_E \ll \omega_s$, the energy deviation and longitudinal co-ordinate damp as:
\begin{eqnarray}
\delta (t) & = & \delta_0 \exp (-\alpha_E t) \sin (\omega_s t - \theta_0), \\
 & & \nonumber \\
z(t) & = & \frac{\alpha_p c}{\omega_s} \delta_0
\exp (-\alpha_E t) \cos (\omega_s t - \theta_0).
\end{eqnarray}
where $\delta_0$ is a constant (the amplitude of the oscillation in $\delta$ at $t = 0$),
and $\theta_0$ is a fixed phase (the phase of the oscillation at $t = 0$).

To find an explicit expression for the damping constant $\alpha_E$, we need to
know how the energy loss per turn $U$ depends on the energy deviation $\delta$.
The total energy lost per turn by a particle is found by integrating the
synchrotron radiation power over one revolution period:
\begin{equation}
U = \oint P_\gamma \, dt.
\end{equation}
To convert this to an integral over the circumference, we should recall that the
path length depends on the energy deviation; so a particle with a higher energy
takes longer to travel around the lattice:
\begin{equation}
dt = \frac{dC}{c} = \frac{1}{c} \left( 1 + \frac{x}{\rho} \right) ds
 = \frac{1}{c} \left( 1 + \frac{\eta_x \delta}{\rho} \right) \, ds.
\end{equation}
Therefore, the radiation energy loss per turn is:
\begin{equation}
U = \frac{1}{c} \oint P_\gamma \left( 1 + \frac{\eta_x \delta}{\rho} \right) \, ds.
\end{equation}
Using (\ref{synchrotronradiationpower}), we find after some algebra:
\begin{equation}
\left. \frac{dU}{dE} \right|_{E = E_0} = j_z \frac{U_0}{E_0},
\end{equation}
where $U_0$ is given by (\ref{energylossperturn}), and the
\emph{longitudinal damping partition number} $j_z$ is:
\begin{equation}
j_z = 2 + \frac{I_4}{I_2}.
\label{longitudinaldampingpartition}
\end{equation}
$I_2$ and $I_4$ are the same synchrotron radiation integrals that we saw
before, in (\ref{eqn_i2}) and (\ref{eqn_i4}).
Finally, we can write the longitudinal damping time:
\begin{equation}
\tau_z = \frac{1}{\alpha_E} = \frac{2}{j_z} \frac{E_0}{U_0} T_0.
\end{equation}

Neglecting coupling, the longitudinal emittance can be given by a similar expression to the horizontal
and vertical emittance:
\begin{equation}
\varepsilon_z = \sqrt{ \langle z^2 \rangle \langle \delta^2 \rangle - \langle z \delta \rangle^2}.
\label{longitudinalemittance}
\end{equation}
Even where dispersion is present, so that the horizontal and longitudinal motion are
coupled, the expression (\ref{longitudinalemittance})
can provide a useful definition of the longitudinal emittance, since the longitudinal variables
usually have a much weaker dependence on the transverse variables, than the transverse
variables have on the longitudinal. Since the amplitudes of the synchrotron oscillations decay
with time constant $\tau_z$, the damping of the longitudinal emittance can be written:
\begin{equation}
\varepsilon_z (t) = \varepsilon_z (t=0) \exp \! \left( -2 \frac{t}{\tau_z} \right). 
\label{longitudinalemittanceevolution}
\end{equation}

It is worth commenting on the fact that the horizontal, vertical and longitudinal
emittances are all damped by synchrotron radiation with exponential damping times
that depend on the beam energy and the rate at which particles lose energy
through synchrotron radiation.  In the case of the horizontal and longitudinal emittances,
there is an additional factor in the expressions for the damping times that depends
on details of the lattice, or, more precisely, on the properties of the dipoles.
The additional factors are given by the damping partition numbers $j_x$ and $j_z$.
From (\ref{horizontaldampingpartition}) and (\ref{longitudinaldampingpartition}), we
see that:
\begin{equation}
j_x + j_z = 3.
\end{equation}
In general, there can also be a vertical damping partition number $j_y$, although
in the simple case we have considered here (of a perfectly planar storage ring)
$j_y = 1$.  A more general analysis would lead to the result:
\begin{equation}
j_x + j_y + j_z = 4,
\end{equation}
which is known as the \emph{Robinson damping theorem} \cite{robinson1958}. 
The significance of this result is that while it is possible (for example, by changing
the field gradient in the dipoles) to `shift' the radiation damping between the
different degrees of freedom, the overall amount of damping is fixed.  In a planar
storage ring, for example, one can reduce the horizontal damping time, but only
at the expense of increasing the longitudinal damping time.

In a typical storage ring, the dispersion in the dipoles is small compared to the
bending radius of the dipoles, that is:
\begin{equation}
\frac{\eta_x}{\rho} \ll 1.
\end{equation}
Then, if there is no quadrupole component in the dipoles (so that $k_1 = 0$
in the dipoles), comparing (\ref{eqn_i2}) and (\ref{eqn_i4}) leads to:
\begin{equation}
\frac{I_4}{I_2} \ll 1,
\end{equation}
in which case:
\begin{eqnarray}
j_x & \approx & 1, \\
j_z & \approx & 2.
\end{eqnarray}
The horizontal damping time is approximately equal to the vertical damping time;
the longitudinal damping time is about half the vertical damping time.
Typical values for the damping times in medium energy synchrotron
light sources are some tens of milliseconds, or a few thousand turns.

\subsection{Quantum excitation}

So far, we have assumed a purely classical model for the radiation, in which
energy can be radiated in arbitrarily small amounts.  From the expressions for the
evolutions of the emittances (\ref{verticalemittanceevolution}),
(\ref{horizontalemittanceevolution}) and (\ref{longitudinalemittanceevolution}),
we see that if radiation was a purely classical process, the emittances would damp
towards zero.  However, quantum effects mean that radiation is emitted in discrete
units (photons).  As we shall see, this induces some `noise' on the beam, known
as \emph{quantum excitation}, the effect of which is to increase the emittance.
The beam in an electron (or positron) storage ring will eventually reach an equilibrium
distribution determined by a balance between the radiation damping and the quantum
excitation.  In the remainder of this section, we shall derive expressions for the rate of
quantum excitation and for the equilibrium emittances in an electron storage ring.

In deriving the equation of motion (\ref{dJxdt}) for the action of a particle
emitting synchrotron radiation, we made the (classical) approximation that in
a time interval $dt$, the momentum $dP$ of the radiation emitted goes to
zero as $dt$ goes to zero.
In reality, emission of radiation is quantized, so we are prevented from taking
the limit $dP \to 0$.
The equation of motion for the action (\ref{rateofchangeofaction}) should then be written:
\begin{equation}
\frac{dJ_x}{dt} = - \frac{w_1}{P_0c} \int_0^\infty \dot{N}(u) \, u \, du+
\frac{w_2}{P_0^2 c^2} \int_0^\infty \dot{N}(u) \, u^2 \, du,
\label{rateofchangeofactionquantum}
\end{equation}
where $\dot{N}(u)$ is the number of photons emitted per unit time in the
energy range from $u$ to $u + du$.
The first term on the right hand side of (\ref{rateofchangeofactionquantum})
just gives the same radiation damping as in the classical approximation;
the second term is an excitation term that we previously neglected.

To find an explicit expression for the rate of change of the action in terms of
the beam and lattice parameters, we need to find expressions for the integrals
$\int \dot{N}(u) \, u\, du$ and $\int \dot{N}(u) \, u^2\, du$.
The required expressions can be found from the spectral distribution of
synchrotron radiation from a dipole magnet, which is another result that
we quote from synchrotron radiation theory.  The spectral distribution of
radiation from a dipole magnet is given by \cite{jackson1998}:
\begin{equation}
\frac{d\mathcal{P}}{d\vartheta} = \frac{9\sqrt{3}}{8\pi} P_\gamma \vartheta \int_\vartheta^\infty K_{5/3} (x) \, dx,
\label{synchrotronradiationspectrum}
\end{equation}
where $d\mathcal{P}/d\vartheta$ is the energy radiated per unit time per
unit frequency range, and $\vartheta = \omega / \omega_c$ is the radiation
frequency $\omega$ divided by the critical frequency $\omega_c$:
\begin{equation}
\omega_c = \frac{3}{2} \frac{\gamma^3 c}{\rho}.
\end{equation}
$P_\gamma$ is the total energy radiated per unit time (\ref{synchrotronradiationpower}), and $K_{5/3}(x)$
is a modified Bessel function.
Since the energy of a photon of frequency $\omega$ is $u = \hbar \omega$,
it follows that:
\begin{equation}
\dot{N}(u) \, du = \frac{1}{\hbar \omega} \frac{d\mathcal{P}}{d\vartheta} \, d\vartheta.
\label{ndotudu}
\end{equation}
Using (\ref{synchrotronradiationspectrum}) and (\ref{ndotudu}), we find:
\begin{equation}
\int_0^\infty \dot{N}(u) \, u \, du = P_\gamma,
\label{integralndotu}
\end{equation}
and:
\begin{equation}
\int_0^\infty \dot{N}(u) \, u^2 \, du = 2 C_q \gamma^2 \frac{E_0}{\rho} P_\gamma.
\label{integralndotusquared}
\end{equation}
$C_q$ is a constant given by:
\begin{equation}
C_q = \frac{55}{32\sqrt{3}} \frac{\hbar}{mc}.
\end{equation}
For electrons (or positrons) $C_q \approx 3.832 \times 10^{-13}\,\mathrm{m}$.

The next step is to substitute for the integrals in (\ref{rateofchangeofactionquantum})
from (\ref{integralndotu}) and (\ref{integralndotusquared}),
substitute for $w_1$ and $w_2$ from (\ref{w1}) and (\ref{w2}), and
average over the circumference of the ring.
This gives an expression for the evolution of the horizontal action (for $x \ll \eta_x$ and $p_x \ll \eta_{px}$):
\begin{equation}
\frac{d\varepsilon_x}{dt} = -\frac{2}{\tau_x} \varepsilon_x + \frac{2}{j_x \tau_x} C_q \gamma^2 \frac{I_5}{I_2},
\label{emittanceevolutionqe}
\end{equation}
where the fifth synchrotron radiation integral $I_5$ is given by:
\begin{equation}
I_5 = \oint \frac{\mathcal{H}_x}{|\rho^3|} \, ds.
\label{radiationintegrali5}
\end{equation}
The $\mathcal{H}$ \emph{function} ($\mathcal{H}_x$) is given by:
\begin{equation}
\mathcal{H}_x = \gamma_x \eta_x^2 + 2 \alpha_x \eta_x \eta_{px}
 + \beta_x \eta_{px}^2.
\end{equation}
The damping time and horizontal damping partition number are given, as
before, by (\ref{horizontaldampingtime1}) and (\ref{horizontaldampingpartition}).
Note that the excitation term is independent of the emittance:
the quantum excitation does not simply modify the damping time,
but leads to a non-zero equilibrium emittance.
The equilibrium emittance $\varepsilon_0$ is determined by the condition:
\begin{equation}
\left. \frac{d\varepsilon_x}{dt} \right|_{\varepsilon_x = \varepsilon_0} = 0,
\end{equation}
and is given by:
\begin{equation}
\varepsilon_0 = C_q \gamma^2 \frac{I_5}{j_x I_2}.
\label{naturalemittance1}
\end{equation}
Note that $\varepsilon_0$ is determined by the beam energy, the lattice
functions (Courant--Snyder parameters and dispersion) in the dipoles, and the bending
radius in the dipoles.  We shall discuss how the design of the lattice affects the value
of $I_5$ (and hence, the equilibrium horizontal emittance) in Section~\ref{section2}.
The equilibrium horizontal emittance (\ref{naturalemittance1}) determined by radiation
is sometimes called the \emph{natural emittance} of the lattice, since
it includes only the most fundamental effects that contribute to the emittance:
radiation damping and quantum excitation.  Other phenomena (such as impedance or
scattering effects) can lead to some increase in the equilibrium emittance actually
achieved in a storage ring, compared to the natural emittance.
Typically, third generation synchrotron light sources have natural emittances of
order of a few nanometres.  With beta functions of a few metres, this implies
horizontal beam sizes of tens of microns (in the absence of dispersion).

In many storage rings, the vertical dispersion in the absence of alignment,
steering and coupling errors is zero, so that $\mathcal{H}_y = 0$.
However, the equilibrium vertical emittance is larger than zero, because the vertical
opening angle of the radiation excites some vertical betatron oscillations.
The fundamental lower limit on the vertical emittance, from the opening
angle of the synchrotron radiation, is given by \cite{raubenheimer1991}:
\begin{equation}
\varepsilon_y = \frac{13}{55} \frac{C_q}{j_y I_2} \oint \frac{\beta_y}{| \rho^3 |}\, ds.
\end{equation}
In most storage rings, this is an extremely small value, typically four orders
of magnitude smaller than the natural (horizontal) emittance.
In practice, the vertical emittance is dominated by magnet alignment errors. 
Storage rings typically operate with a vertical emittance that is of order 1\%
of the horizontal emittance, but many can achieve emittance ratios
somewhat smaller than this.  We shall discuss the vertical emittance in more
detail in Section~\ref{section3}.

\begin{figure}[t]
\begin{center}
\includegraphics[width=0.5\textwidth]{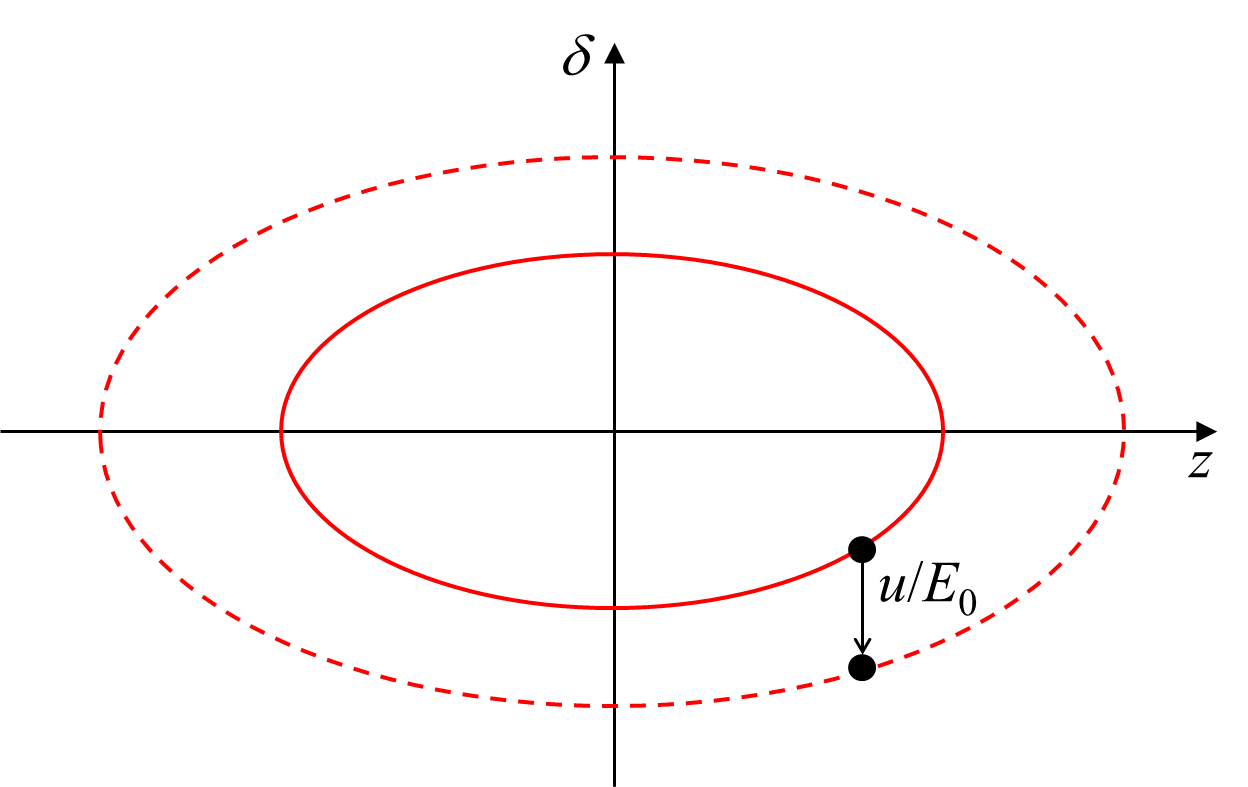}
\caption{Change in longitudinal phase space variables for a particle emitting a
photon carrying energy $u$.  As a result of the photon emission, there is a change
in amplitude of the synchrotron oscillations (represented by the ellipses) performed
by the particle as it moves around the storage ring.}
\label{figlongitudinalphasespace}
\end{center}
\end{figure}

Quantum effects excite longitudinal emittance as well as transverse emittance.
Consider a particle with longitudinal co-ordinate $z$ and energy deviation $\delta$,
which emits a photon of energy $u$ (see Fig.~\ref{figlongitudinalphasespace}).  The
co-ordinate and energy deviation after emission of the photon are given by:
\begin{eqnarray}
\delta^\prime & = & \delta_0^\prime \sin \theta^\prime =
\delta_0 \sin \theta - \frac{u}{E_0}, \\
z^\prime & = & \frac{\alpha_p c}{\omega_s} \delta_0^\prime \cos \theta^\prime =
 \frac{\alpha_p c}{\omega_s} \delta_0 \cos \theta.
\end{eqnarray}
Therefore:
\begin{equation}
\delta_0^{\prime 2} = \delta_0^2 - 2\delta_0 \frac{u}{E_0} \sin \theta + \frac{u^2}{E_0^2}.
\end{equation}
Averaging over the bunch gives:
\begin{equation}
\Delta \sigma_\delta^2 = \frac{\langle u^2 \rangle}{2E_0^2},
\end{equation}
where:
\begin{equation}
\sigma_\delta^2 = \langle \delta^2 \rangle = \frac{1}{2} \langle \delta_0^2 \rangle.
\end{equation}
Including radiation damping, the energy spread evolves as:
\begin{equation}
\frac{d\sigma_\delta^2}{dt} = \frac{1}{2E_0^2}
\frac{1}{C_0} \oint dC \, \int_0^\infty du \, \dot{N}(u) u^2 - \frac{2}{\tau_z} \sigma_\delta^2,
\end{equation}
where we have averaged the radiation effects around the ring by integrating over the
circumference.
Using (\ref{integralndotusquared}) for $\int \dot{N}(u) u^2 \, du$, we find:
\begin{equation}
\frac{d\sigma_\delta^2}{dt} = C_q \gamma^2 \frac{2}{j_z \tau_z} \frac{I_3}{I_2}
- \frac{2}{\tau_z} \sigma_\delta^2.
\end{equation}
The equilibrium energy spread is given by $d\sigma_\delta^2 /dt = 0$:
\begin{equation}
\sigma_{\delta 0}^2 = C_q \gamma^2 \frac{I_3}{j_z I_2},
\label{naturalenergyspreadmeansquare}
\end{equation}
where the third synchrotron radiation integral $I_3$ is defined:
\begin{equation}
I_3 = \oint \frac{1}{| \rho^3 |} \, ds.
\end{equation}
The equilibrium energy spread $\sigma_{\delta 0}$ determined by radiation effects is
often referred to as the \emph{natural energy spread}, since collective effects can
often lead to an increase in the energy spread with increasing bunch charge.  Note
that the natural energy spread is determined essentially by the beam energy and by
the bending radii of the dipoles; rather counterintuitively, it does not depend on the
rf parameters (either the voltage or the frequency).  On the other hand, the bunch
length does have a dependence on the rf.  The ratio of the bunch length $\sigma_z$
to the energy spread $\sigma_\delta$ in a \emph{matched distribution} (i.e.~a
distribution that is unchanged after one complete revolution around the ring) can be
determined from the shape of the ellipse in longitudinal phase space followed by a
particle obeying the longitudinal equations of motion (\ref{longitudinaleom1}) and
(\ref{longitudinaleom2}).  Neglecting radiation effects (which can be assumed to be
small) the result is:
\begin{equation}
\sigma_z = \frac{\alpha_p c}{\omega_s} \sigma_\delta.
\end{equation}
We can increase the synchrotron frequency $\omega_s$, and hence reduce the
bunch length, by increasing the rf voltage, or by increasing the rf frequency.

\subsection{Summary of radiation damping and quantum excitation}

To summarise, including the effects of radiation damping and quantum excitation, the
emittances (in each of the three degrees of freedom) evolve with time as:
\begin{equation}
\varepsilon (t) = \varepsilon(t=0) \exp \left( -2\frac{t}{\tau} \right)
+ \varepsilon(t = \infty) \left[ 1 - \exp \left( -2\frac{t}{\tau} \right) \right] ,
\end{equation}
where $\varepsilon(t=0)$ is the initial emittance (for example, of a beam as it is injected into
the storage ring), and $\varepsilon(t=\infty)$ is the equilibrium emittance determined by
the balance between radiation damping and quantum excitation.
The damping times are given by:
\begin{equation}
j_x \tau_x = j_y \tau_y = j_z \tau_z = 2\frac{E_0}{U_0} T_0,
\end{equation}
where the damping partition numbers are given by:
\begin{equation}
j_x = 1 - \frac{I_4}{I_2}, \qquad
j_y = 1, \qquad
j_z = 2 + \frac{I_4}{I_2}.
\end{equation}
The energy loss per turn is given by:
\begin{equation}
U_0 = \frac{C_\gamma}{2\pi} E_0^4 I_2,
\end{equation}
where for electrons (or positrons) $C_\gamma \approx 8.846\times 10^{-5} \textrm{ m/GeV}^3.$
The natural emittance is:
\begin{equation}
\varepsilon_0 = C_q \gamma^2 \frac{I_5}{j_x I_2},
\end{equation}
where for electrons (or positrons) $C_q \approx 3.832\times 10^{-13} \textrm{ m}.$
The natural rms energy spread and bunch length are given by:
\begin{eqnarray}
\sigma_\delta^2 & = & C_q \gamma^2 \frac{I_3}{j_z I_2}, \\
\sigma_z & = & \frac{\alpha_p c}{\omega_s} \sigma_\delta.
\end{eqnarray}
The momentum compaction factor is:
\begin{equation}
\alpha_p = \frac{I_1}{C_0}.
\end{equation}
The synchrotron frequency and synchronous phase are given by:
\begin{eqnarray}
\omega_s^2 & = & -\frac{qV_\textrm{rf}}{E_0} \frac{\omega_\textrm{rf}}{T_0} \alpha_p \cos \phi_s, \\
\sin \phi_s & = & \frac{U_0}{qV_\textrm{rf}}.
\end{eqnarray}
Finally, the synchrotron radiation integrals are:
\begin{eqnarray}
I_1 & = & \oint \frac{\eta_x}{\rho} \, ds, \\
I_2 & = & \oint \frac{1}{\rho^2} \, ds, \\
I_3 & = & \oint \frac{1}{| \rho |^3} \, ds, \\
I_4 & = & \oint \frac{\eta_x}{\rho} \left( \frac{1}{\rho^2} + 2k_1 \right) \, ds, 
\qquad k_1 = \frac{e}{P_0} \frac{\partial B_y}{\partial x}, \\
I_5 & = & \oint \frac{\mathcal{H}_x}{| \rho |^3} \, ds, \qquad
\mathcal{H}_x = \gamma_x \eta_x^2 + 2\alpha_x \eta_x \eta_{px} + \beta_x \eta_{px}^2.
\end{eqnarray}

\section{Equilibrium emittance and storage ring lattice design\label{section2}}

In this section, we shall derive expressions for the natural emittance in four types of lattices:
FODO, double bend achromat (DBA), multi-bend achromat (including the triple bend achromat)
and theoretical minimum emittance (TME) lattices.
We shall also consider how the emittance of an achromat may be reduced by
`detuning' the lattice from the strict achromat conditions.

Recall that the natural emittance in a storage ring is
given by (\ref{naturalemittance1}):
\begin{equation}
\varepsilon_0 = C_q \gamma^2 \frac{I_5}{j_x I_2},
\end{equation}
where $C_q$ is a physical constant, $\gamma$ is the relativistic factor,
$j_x$ is the horizontal damping partition number, and $I_5$ and $I_2$
are synchrotron radiation integrals.
Note that $j_x$, $I_5$ and $I_2$ are all fixed by the layout of the lattice
and the optics, and are
independent of the beam energy.
In most storage rings, if the bends have no quadrupole component, the
damping partition number $j_x \approx 1$.
In that case, to find the natural emittance we just need to evaluate the
two synchrotron radiation integrals $I_2$ and $I_5$.
If we know the strength and length of all the dipoles in the lattice, it is
straightforward to calculate $I_2$.
For example, if all the bends are identical, then in a complete ring
(total bending angle = 2$\pi$):
\begin{equation}
I_2 = \oint \frac{1}{\rho^2} ds
 = \oint \frac{B}{(B\rho)} \frac{ds}{\rho}
 = \frac{2\pi B}{(B\rho)}
 \approx 2\pi \frac{c B}{E/q},
\end{equation}
where $E$ is the beam energy, and $q$ is the particle charge.
Evaluating $I_5$ is more complicated: it depends on the lattice functions.

\subsection{FODO lattice}

\begin{figure}[t]
\begin{center}
\includegraphics[width=0.7\textwidth]{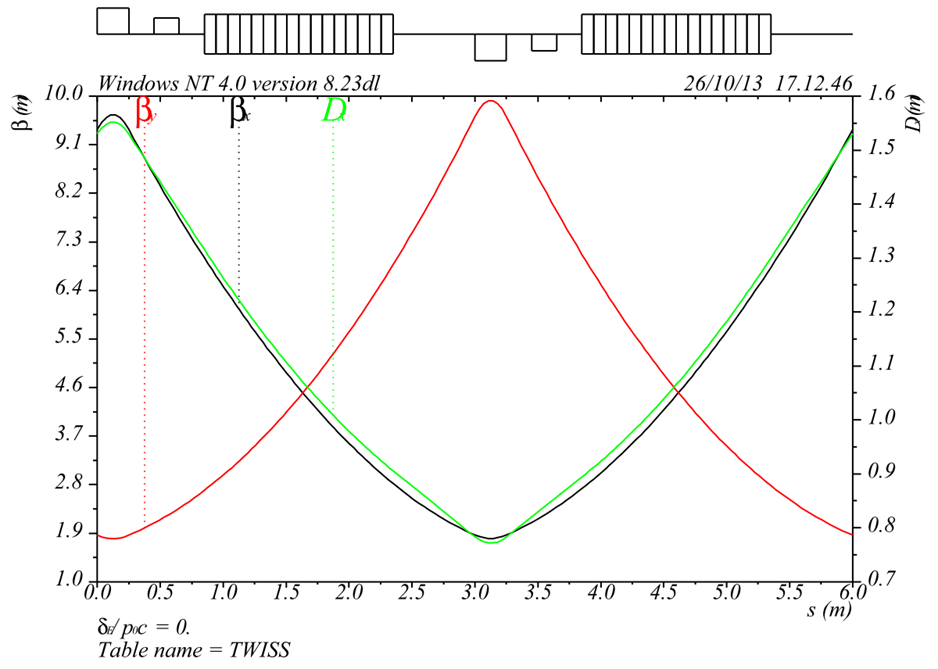}
\includegraphics[width=0.7\textwidth]{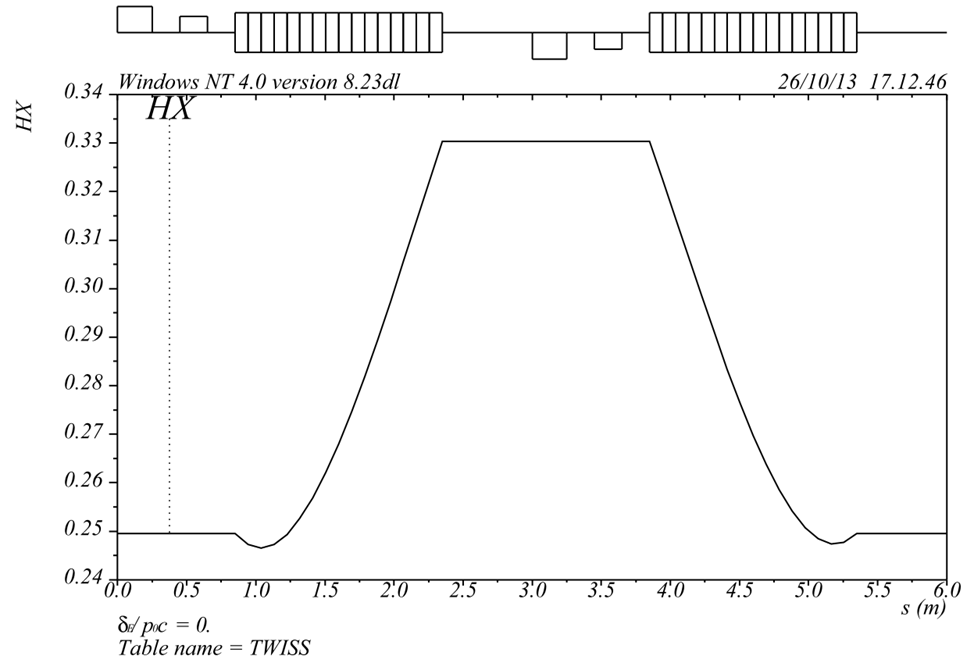}
\caption{Lattice functions in a FODO cell.
Top:  Courant--Snyder parameters and dispersion.
Bottom: $\mathcal{H}$ function.
In this case, the phase advance is 90$^\circ$, the dipoles are 1.5\,m long
and have bending angle 2$\pi$/32.  Notice that the value of the
$\mathcal{H}$ function is constant except in the dipoles: this is a general
property of this function.}
\label{figfodolatticefunctions}
\end{center}
\end{figure}

Let us consider the case of a FODO lattice.  The lattice functions in a typical FODO cell
are shown in Fig.~\ref{figfodolatticefunctions}.  To simplify the system,
we use the following approximations.  First, we assume that
the quadrupoles can be represented by thin lenses.
Second, we assume that the space between the quadrupoles is completely
filled by the dipoles.  This is clearly not a realistic assumption, but it does
allow us to derive some useful (and reasonably accurate) formulae.
With these approximations, the lattice functions (Courant--Snyder parameters and dispersion)
are completely determined by the focal length $f$ of the quadrupoles
and the bending radius $\rho$ and length $L$ of the dipoles, and can
be calculated using standard techniques.

Suppose that $R_\textrm{cell}$ is the transfer matrix for the horizontal motion in one
complete periodic cell of a lattice.  $R_\textrm{cell}$ may be constructed by multiplying the
transfer matrices $R$ for individual components in the beam line.  For example,
for a thin quadrupole of focal length $f$:
\begin{equation}
R_\textrm{quad} = \left( \begin{array}{cc}
1 & 0 \\
-1/f & 1
\end{array} \right).
\end{equation}
For a dipole of bending radius $\rho$ and length $L$, the transfer matrix is:
\begin{equation}
R_\textrm{dip} = \left( \begin{array}{cc}
\cos \frac{L}{\rho} & \rho \sin \frac{L}{\rho} \\
-\frac{1}{\rho} \sin \frac{L}{\rho} & \cos \frac{L}{\rho}
\end{array} \right).
\end{equation}
The Courant--Snyder parameters at any point in the beam line can be found
first by multiplying the transfer matrices $R$ for the individual components
to give the transfer matrix $R_\textrm{cell}$ for the periodic cell starting from
the chosen point, and then writing the complete transfer matrix in the form:
\begin{equation}
R_\textrm{cell} = \left( \begin{array}{cc}
\cos \mu_x + \alpha_x \sin \mu_x & \beta_x \sin \mu_x \\
-\gamma_x \sin \mu_x & \cos \mu_x - \alpha_x \sin \mu_x
\end{array} \right),
\end{equation}
where $\mu_x$ is the phase advance.
The dispersion describes the periodic trajectory of an (off-energy) particle
through a periodic cell, and can be found at any point by solving the condition:
\begin{equation}
\left( \begin{array}{c}
\eta_x \\ \eta_{px}
\end{array} \right) = R_\textrm{cell}^\eta
\left( \begin{array}{c}
\eta_x \\ \eta_{px}
\end{array} \right) + 
d_\textrm{cell}^\eta,
\end{equation}
where $R_\textrm{cell}^\eta$ is a matrix representing the first order terms in the map
(for a complete cell) for the dispersion, and $d_\textrm{cell}^\eta$ is a vector representing
the zeroth order terms.  The map for a complete cell is found, as usual,
by composing the maps for individual elements.  For a quadrupole,
the map for the dispersion is the same as the map for the dynamical
variables; for a dipole, there are additional zeroth order terms:
\begin{equation}
\left( \begin{array}{c}
\eta_x \\ \eta_{px}
\end{array} \right)_{\!\! s_0+L} = R_\textrm{dip}
\left( \begin{array}{c}
\eta_x \\ \eta_{px}
\end{array} \right)_{\!\! s_0} + 
\left( \begin{array}{c}
\rho (1 - \cos \frac{L}{\rho}) \\ \sin \frac{L}{\rho}
\end{array} \right).
\label{dispersionevolutionindipole}
\end{equation}

Using the above results, we find that
in terms of $f$, $\rho$ and $L$, the horizontal beta function at the
horizontally focusing quadrupole in a FODO cell is given by:
\begin{equation}
\beta_x = \frac{
4f\rho \sin \theta (2f \cos \theta + \rho \sin \theta)
}{
\sqrt{16f^4 - [\rho^2 - (4f^2 + \rho^2)\cos 2\theta]^2}
},
\end{equation}
where $\theta = L/\rho$ is the bending angle of a single dipole.
The dispersion at a horizontally focusing quadrupole is given by:
\begin{equation}
\eta_x = \frac{
2f \rho (2f + \rho \tan \frac{\theta}{2})
}
{
4f^2 + \rho^2
}.
\end{equation}
By symmetry, at the centre of a quadrupole, $\alpha_x = \eta_{px} = 0$.
Given the lattice functions at any point in the lattice, we can evolve the functions
through the lattice, using the transfer matrices $R$.
For the Courant--Snyder parameters:
\begin{equation}
A(s_1) = R A(s_0) R\transpose,
\end{equation}
where $R = R(s_1; s_0)$ is the transfer matrix from $s_0$ to $s_1$,
$R\transpose$ is the transpose of $R$, and:
\begin{equation}
A = \left( \begin{array}{cc}
 \beta_x & -\alpha_x \\
-\alpha_x & \gamma_x
\end{array} \right).
\end{equation}
The dispersion can be evolved (over a distance $L$, with constant
bending radius $\rho$) using (\ref{dispersionevolutionindipole}).

\begin{figure}[t]
\begin{center}
\includegraphics[width=0.7\textwidth]{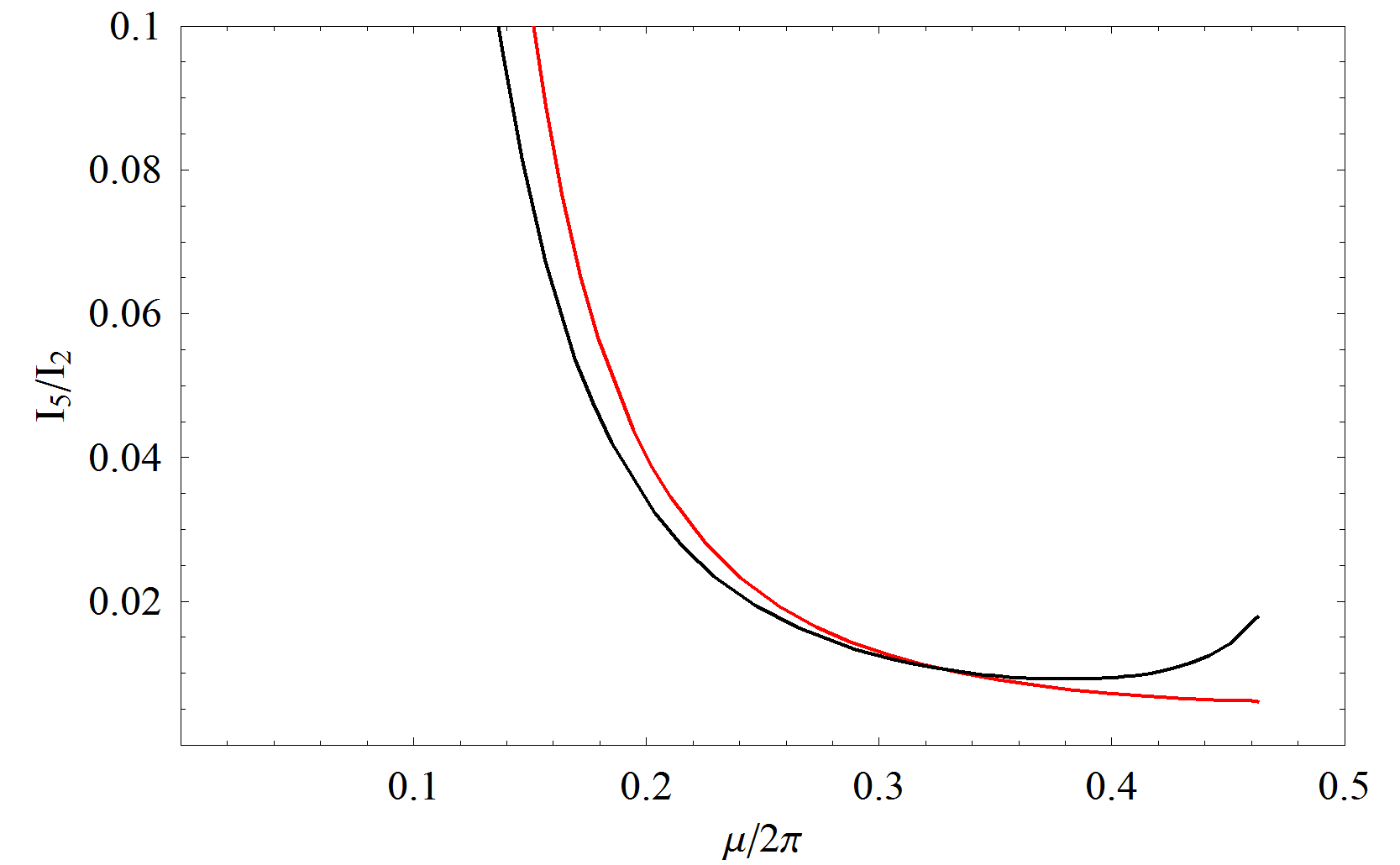}
\caption{Ratio of synchrotron radiation integrals $I_5/I_2$ in a FODO cell,
as a function of the phase advance.  The black line shows the exact value,
while the red line shows the value calculated using the approximation
(\ref{fodoi5di2approximation}).}
\label{figfodoi5di2}
\end{center}
\end{figure}

We now have all the information we need to find an expression for $I_5$
in the FODO cell.
However, the algebra is rather formidable.  The result is most easily
expressed as a power series in the dipole bending angle, $\theta$:
\begin{equation}
\frac{I_5}{I_2} = \left( 4 + \frac{\rho^2}{f^2} \right)^{\!\!-\frac{3}{2}}
\left(
8 - \frac{\rho^2}{2f^2}\theta^2 + O(\theta^4)
\right).
\label{fodoi5di2}
\end{equation}
For small $\theta$, the expression for $I_5/I_2$ can be written:
\begin{equation}
\frac{I_5}{I_2} \approx \left( 1 - \frac{\rho^2}{16f^2} \theta^2 \right)
\left( 1 + \frac{\rho^2}{4f^2} \right)^{\!\!-\frac{3}{2}}
= \left( 1 - \frac{L^2}{16f^2} \right)
\left( 1 + \frac{\rho^2}{4f^2} \right)^{\!\!-\frac{3}{2}}.
\end{equation}
This can be further simplified if $\rho \gg 2f$ (which is often the case):
\begin{equation}
\frac{I_5}{I_2} \approx \left( 1 - \frac{L^2}{16f^2} \right)
\frac{8f^3}{\rho^3},
\label{fodoi5di2approximation}
\end{equation}
and still further simplified if $4f \gg L$ (which is less often the case):
\begin{equation}
\frac{I_5}{I_2} \approx \frac{8f^3}{\rho^3}.
\end{equation}
The ratio $I_5/I_2$ is plotted for a FODO cell as a function of the phase
advance in Fig.~\ref{figfodoi5di2}.
Making the approximation $j_x \approx 1$ (since we assume that there is no quadrupole
component in the dipole), and writing $\rho = L/\theta$, we have:
\begin{equation}
\varepsilon_0 \approx C_q \gamma^2 \left( \frac{2f}{L} \right)^{\!\!3} \theta^3.
\label{fodoemittance1}
\end{equation}

Notice how the emittance scales with the beam and lattice parameters.
The emittance is proportional to the square of the energy
and to the cube of the bending angle.
Increasing the number of cells in a complete circular lattice reduces the
bending angle of each dipole, and reduces the emittance.
The emittance is proportional to the cube of the quadrupole focal
length: stronger focusing results in lower emittance.
Finally, the emittance is inversely proportional to the cube of the cell length.

The phase advance in a FODO cell is given by:
\begin{equation}
\cos \mu_x = 1 - \frac{L^2}{2f^2}.
\end{equation}
This means that a stable lattice must have:
\begin{equation}
\frac{f}{L} \ge \frac{1}{2}.
\end{equation}
In the limiting case, $\mu_x = 180^\circ$, and $f$ has the minimum value
$f = L/2$.  Using the approximation (\ref{fodoemittance1}) gives:
\begin{equation}
\varepsilon_0 \approx C_q \gamma^2 \left( \frac{2f}{L} \right)^{\!\!3} \theta^3,
\nonumber
\end{equation}
and so the minimum emittance in a FODO lattice is expected to be:
\begin{equation}
\varepsilon_{0,\textrm{FODO,min}} \approx C_q \gamma^2 \theta^3.
\end{equation}
However, as we increase the focusing strength, the approximations we used
to obtain the simple expression for $\varepsilon_0$ start to break down.
From the exact formula for $I_5/I_2$ as a function of the phase advance,
we find (by numerical means) that there is a minimum in the natural emittance
at $\mu_x \approx 137^\circ \approx 0.38\times 2\pi\,$rad
(see Fig.~\ref{figfodoi5di2}).  The minimum value of the natural emittance
in a FODO lattice is given by:
\begin{equation}
\varepsilon_{0,\textrm{FODO,min}} \approx 1.2 C_q \gamma^2 \theta^3.
\label{minimumemittancefodo}
\end{equation}

As an example, consider a storage ring with 16 FODO cells (32 dipoles),
90$^\circ$ phase advance per cell ($f = L/\sqrt{2}$), and with a stored
beam energy of 2\,GeV.  Using (\ref{fodoemittance1}) we estimate that
such a ring would have a natural emittance of around 125\,nm.  Many
modern applications (including synchrotron light sources) demand emittances
smaller than this by one or two orders of magnitude.  This raises the
question of how we might design a lattice with a smaller natural emittance.
Looking at the lattice functions in a FODO lattice
(Fig.~\ref{figfodolatticefunctions}) provides a clue.
The dispersion function, which is directly related to the effect of quantum
excitation on the horizontal emittance, is non-zero throughout the cell.
If we can design a lattice where the dispersion vanishes at the entrance
of a dipole, then we might hope to reduce the average value of the
$\mathcal{H}$ function in the dipoles, thereby reducing $I_5$ and the
value of the natural emittance.  It is indeed possible to design a cell with two
dipoles, in which the dispersion vanishes at the entrance of the first dipole
and at the exit of the second dipole: such a cell is known as a
\emph{Chasman--Green cell} \cite{chasmangreen1975}, or a double bend
achromat (DBA).

\subsection{Double bend achromat lattice}

To calculate the natural emittance in a DBA lattice, let us begin by considering
the conditions for zero dispersion at the start and the exit of a unit cell.
Assume that the dispersion is zero at the start of the cell.  We place a
quadrupole midway between the dipoles, to reverse the gradient of the
dispersion.  By symmetry, the dispersion at the exit of the cell will then also
be zero.  In the thin lens approximation, the required strength of the
quadrupole between the dipoles can be determined from:
\begin{equation}
\left( \begin{array}{cc}
 1 & 0 \\
-1/f & 1
\end{array} \right) 
\left( \begin{array}{c}
\eta_x \\
\eta_{px}
\end{array} \right) = 
\left( \begin{array}{c}
\eta_x \\
\eta_{px} - \frac{\eta_x}{f}
\end{array} \right) = 
\left( \begin{array}{c}
\eta_x \\
-\eta_{px}
\end{array} \right).
\end{equation}
Hence the central quadrupole must have focal length:
\begin{equation}
f = \frac{\eta_x}{2\eta_{px}}.
\end{equation}
The actual value of the dispersion (and its gradient) is determined by the
dipole bending angle $\theta$, the bending radius $\rho$, and the drift
length $L_\textrm{drift}$:
\begin{eqnarray}
\eta_x & = & \rho (1 - \cos \theta ) + L_\textrm{drift} \sin \theta, \\
\eta_{px} & = & \sin \theta.
\end{eqnarray}

\begin{figure}[t]
\begin{center}
\includegraphics[width=0.7\textwidth]{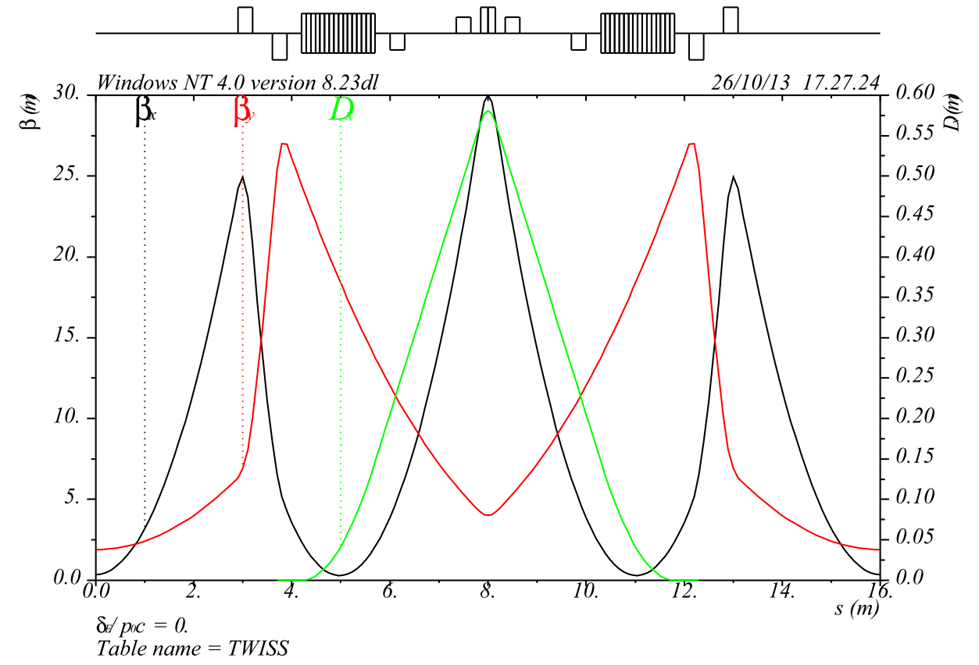}
\includegraphics[width=0.7\textwidth]{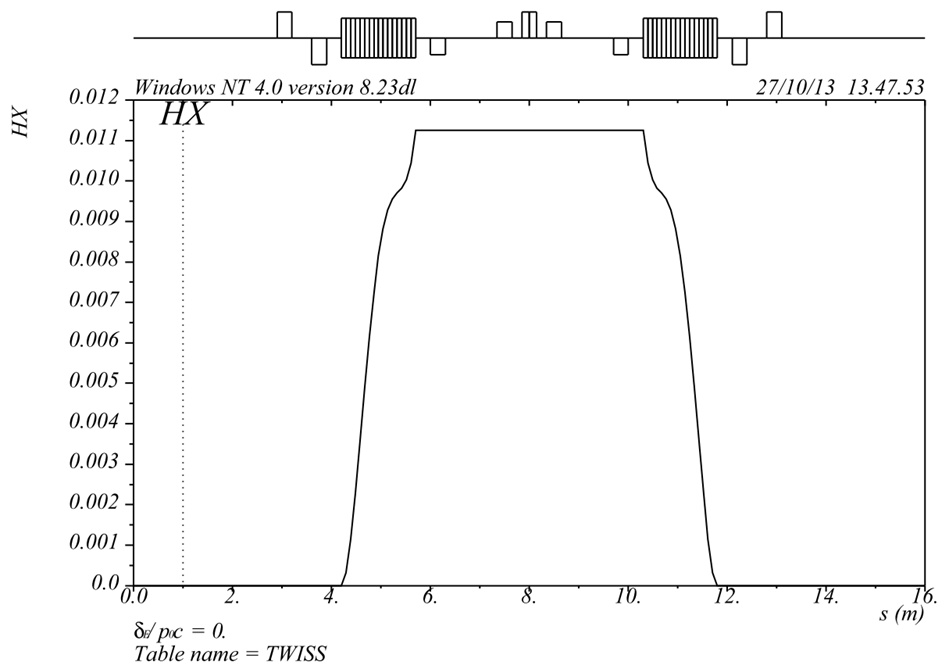}
\caption{Lattice functions in a DBA cell.
Top: Courant--Snyder parameters and dispersion.
Bottom: $\mathcal{H}$ function.
The horizontal beta and alpha functions at the entrance of the first dipole
have values $\beta_x=2.08\,$m and $\alpha_x = 2.47$.  These are
different from the `ideal' values for low emittance in this case, of $\beta_x=2.33\,$m
and $\alpha_x = 3.87$.  The lattice functions are detuned from their ideal values
in order to satisfy a range of constraints (such as maximum values of the beta
functions, magnet strengths and chromaticity).  The detuning results in this
case in an increase in the natural emittance by a factor of 1.8.}
\label{figdbalatticefunctions}
\end{center}
\end{figure}

To complete the DBA cell, we need to include some additional quadrupoles
in the zero-dispersion region to control the horizontal and vertical beta functions.
To correct the chromaticity, sextupoles are included between the dipoles, where
the dispersion is non-zero.  The lattice functions in an example DBA cell are
shown in Fig.~\ref{figdbalatticefunctions}.  To get some idea whether this style
of lattice likely to have a lower natural emittance than a FODO lattice, we can
inspect the $\mathcal{H}$ function.  Comparing Figs.~\ref{figfodolatticefunctions}
and \ref{figdbalatticefunctions},
we see that the $\mathcal{H}$ function is much smaller in the DBA lattice
than in the FODO lattice.  Note that we use the same dipoles (bending angle
and length) in both cases.

Let us calculate the minimum natural emittance of a DBA lattice, for given
bending radius $\rho$ and bending angle $\theta$ in the dipoles.
To do this, we need to calculate the minimum value of:
\begin{equation}
I_5 = \int_0^L \frac{\mathcal{H}_x}{\rho^3}\, ds
\end{equation}
in one dipole (of length $L$), subject to the constraints:
\begin{equation}
\eta_{x,0} = \eta_{px,0} = 0,
\label{dbaconstraints}
\end{equation}
where $\eta_{x,0}$ and $\eta_{px,0}$ are the dispersion and gradient of
the dispersion at the entrance of the dipole.
We know how the dispersion and the Courant--Snyder parameters evolve through the
dipole, so we can calculate $I_5$ for one dipole, for given initial values of the
Courant--Snyder parameters $\alpha_{x,0}$ and $\beta_{x,0}$.
Then, we have to minimise the value of $I_5$ with respect to
$\alpha_{x,0}$ and $\beta_{x,0}$.
Again, the algebra is rather formidable, and the full expression for $I_5$ is
not especially enlightening:
therefore, we just quote the significant results.
We find that, for given $\rho$ and $\theta$ and with the constraints (\ref{dbaconstraints})
the minimum value of $I_5$ is given by:
\begin{equation}
I_{5,\textrm{min}} = \frac{1}{4\sqrt{15}} \frac{\theta^4}{\rho} + O(\theta^6).
\end{equation}
This minimum occurs for values of the Courant--Snyder parameters at the entrance to
the dipole given by:
\begin{eqnarray}
\beta_{x,0} & = & \sqrt{\frac{12}{5}} L + O(\theta^3), \\
\alpha_{x,0} & = & \sqrt{15} + O(\theta^2),
\end{eqnarray}
where $L = \rho \theta$ is the length of a dipole.
Since we know that $I_2$ in a single dipole is given by:
\begin{equation}
I_2 = \int_0^L \frac{1}{\rho^2} \, ds = \frac{\theta}{\rho},
\end{equation}
we can now write down an expression for the minimum emittance in a DBA
lattice:
\begin{equation}
\varepsilon_{0,\textrm{DBA,min}} = C_q \gamma^2 \frac{I_{5,\textrm{min}}}{j_x I_2}
\approx \frac{1}{4\sqrt{15}} C_q \gamma^2 \theta^3.
\label{minimumemittancedba}
\end{equation}
The approximation is valid for small $\theta$.  Note that we have again
assumed that, since there is no quadrupole component in the dipole, $j_x \approx 1$.

Compare the expression (\ref{minimumemittancedba}) for the minimum emittance in
a DBA lattice, with the expression (\ref{minimumemittancefodo}) for the minimum
emittance in a FODO lattice. 
We see that in both cases (FODO and DBA), the emittance scales with the
square of the beam energy, and with the cube of the bending angle.
However, the emittance in a DBA lattice is smaller than that in a FODO
lattice (for given energy and dipole bending angle) by a factor
$4\sqrt{15} \approx 15.5$.

This is a significant improvement; however, there is still the possibility of
reducing the natural emittance (for a given beam energy and number of cells)
even further.
For a DBA lattice, we imposed constraints (\ref{dbaconstraints}) on the dispersion
at the entrance of the first dipole in a lattice cell.
To reach a lower emittance, we can consider relaxing these constraints.

\subsection{Theoretical minimum emittance lattice}

To derive the conditions for a \emph{theoretical minimum emittance} (TME)
lattice, we write down an expression for:
\begin{equation}
I_5 = \int_0^L \frac{\mathcal{H}_x}{\rho} \, ds,
\end{equation}
with \emph{arbitrary} dispersion $\eta_{x,0}$, $\eta_{px,0}$ and Courant--Snyder
parameters $\alpha_{x,0}$ and $\beta_{x,0}$ in a dipole with given bending
radius $\rho$ and angle $\theta$ (and length $L = \rho \theta$).
Then, we minimise $I_5$ with respect to $\eta_{x,0}$, $\eta_{px,0}$,
$\alpha_{x,0}$ and $\beta_{x,0}$.
The result is \cite{leeteng1991}:
\begin{equation}
\varepsilon_{0,\textrm{TME,min}} \approx
\frac{1}{12\sqrt{15}} C_q \gamma^2 \theta^3.
\end{equation}
The minimum emittance is obtained with dispersion at the entrance to the
dipole given by:
\begin{eqnarray}
\eta_{x,0} & = & \frac{1}{6}L\theta + O(\theta^3), \\
\eta_{px,0} & = & -\frac{\theta}{2} + O(\theta^3),
\end{eqnarray}
and with Courant--Snyder functions at the entrance:
\begin{eqnarray}
\beta_{x,0} & = & \frac{8}{\sqrt{15}}L + O(\theta^2), \\
\alpha_{x,0} & = & \sqrt{15} + O(\theta^2).
\end{eqnarray}
The dispersion and beta function reach minimum values in the centre of the
dipole:
\begin{eqnarray}
\eta_{x,\textrm{min}} & = & \rho \left( 1 - \frac{2}{\theta} \sin \!\left(\frac{\theta}{2}\right) \right)
= \frac{L\theta}{24} + O(\theta^4), \\
\beta_{x,\textrm{min}} & = & \frac{L}{2\sqrt{15}} + O(\theta^2).
\end{eqnarray}
By symmetry, we can consider a single TME cell to contain a single dipole,
rather than a pair of dipoles as was necessary for the DBA cell.
Outside the dipole, the dispersion is relatively large.  This is not ideal for
a light source, since insertion devices at locations with large dispersion will
blow up the emittance.  If insertion devices are required, then it is
possible to break the symmetry of the lattice to include zero-dispersion
straights: for example, the ring could have a race-track footprint, with
arcs constructed from TME cells.

\begin{figure}[t]
\begin{center}
\includegraphics[width=0.7\textwidth]{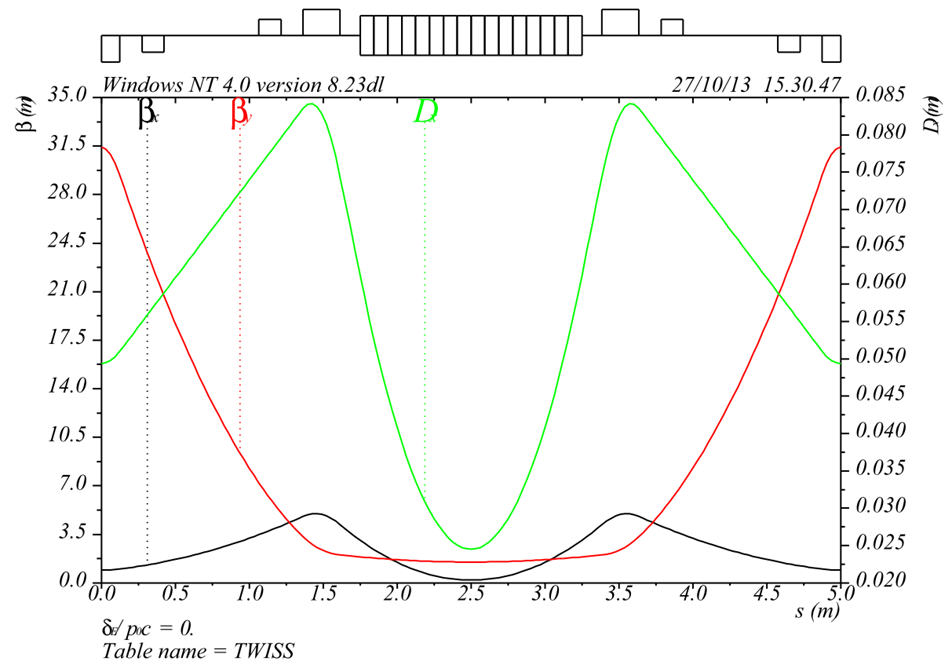}
\includegraphics[width=0.7\textwidth]{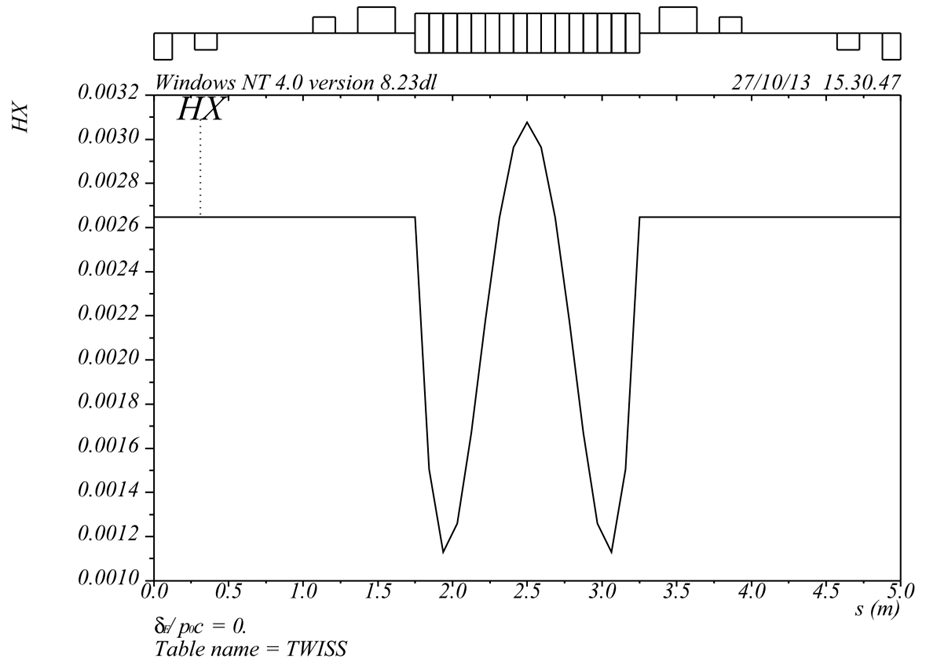}
\caption{Lattice functions in a TME cell.
Top: Courant--Snyder parameters and dispersion.
Bottom: $\mathcal{H}$ function.
The horizontal beta function and dispersion match the `ideal' values
for low emittance.}
\label{figtmelatticefunctions}
\end{center}
\end{figure}

Examples of the lattice functions (and $\mathcal{H}$ function) in a TME
cell are shown in Fig.~\ref{figtmelatticefunctions}.
Note that the $\mathcal{H}$ function in the dipole in the TME cell is significantly
lower than for FODO or DBA cells using similar dipoles
(Figs.~\ref{figfodolatticefunctions} and \ref{figdbalatticefunctions}).

\subsection{Practical constraints on lattice optics}

The results we have derived for the natural emittance in FODO, DBA and TME
lattices have been for `ideal' lattices that perfectly
achieve the stated conditions in each case.
In practice, lattices rarely, if ever, achieve the ideal conditions.  In particular,
the beta function in an achromat is usually not optimal for low emittance; and
it is difficult to tune the dispersion for the ideal TME conditions.
The main reasons for this are: first, 
beam dynamics issues (relating, for example, to nonlinear dynamics and collective effects)
often impose a variety of strong constraints on the design;
and second, optimising the lattice functions while respecting all the various constraints can
require complex configurations of quadrupoles.
A particularly challenging constraint on design of a low-emittance lattice is
the dynamic aperture.
Storage rings require a large dynamic aperture in order to achieve good
injection efficiency and good beam lifetime.
However, low emittance lattices generally need low dispersion and beta
functions, and hence require strong quadrupoles.  As a result, the
chromaticity can be large, and must be corrected using strong sextupoles.
Strong sextupoles lead to highly nonlinear motion and a limited dynamic
aperture: the trajectories of particles at even quite modest betatron
amplitudes or energy deviations can become unstable, resulting in short
beam lifetime.

Lattices composed of DBA cells have been a popular choice for third
generation synchrotron light sources.
The DBA structure provides a lower natural emittance than a FODO
lattice with the same number of dipoles, while
the long, dispersion-free straight sections provide ideal locations for
insertion devices such as undulators and wigglers.
If an insertion device, such as an undulator or wiggler, is incorporated in a
storage ring at a location with large dispersion, then the dipole fields in the
device can make a significant contribution to the quantum excitation ($I_5$).
As a result, the insertion device can lead to an increase in the natural
emittance of the storage ring.
By using a DBA lattice, dispersion-free straights are naturally provided, in which
undulators and wigglers can be located without blowing up the natural emittance.
However, there is some tolerance.  In many cases, it is possible to detune
the lattice from the strict DBA conditions, thereby allowing some reduction in
natural emittance at the cost of some dispersion in the straights.
The insertion devices will then contribute to the quantum excitation; but
depending on the lattice and the insertion devices, there may still be a net
benefit.  Some light sources that were originally designed with zero-dispersion
straights take advantage of tuning flexibility to operate with non-zero dispersion
in the straights (see, for example, \cite{ropert1995}).
This provides a lower natural emittance, and better output for users.


\subsection{Multi-bend achromats}


There are of course many options for the design of a storage ring lattice, beyond
the FODO, DBA and TME cells we have discussed so far.  For example, it is possible
to combine the DBA and TME lattices, constructing an arc cell consisting of more than
two dipoles.  The dipoles at either end of the cell have zero dispersion (and gradient
of the dispersion) at their outside faces, thus satisfying the achromat condition.
Since the lattice functions are different in the central dipoles compared to the end
dipoles, we have additional degrees of freedom we can use to minimise the quantum
excitation.  The result is a \emph{multi-bend achromat} (MBA) that combines the
benefits of a DBA lattice (with long straights providing good locations for
insertion devices) and a TME lattice (providing the possibility of achieving
lower emittance than in a DBA).

In a MBA, it is possible to have cases where the end dipoles and central dipoles differ in
the bend angle (i.e.~length of dipole), and/or the bend radius (i.e.~strength of dipole).
For simplicity, let us consider the case where the dipoles all have the same
bending radius (i.e.~they all have the same field strength), but they vary in
length.  Assume that each arc cell has a fixed number $M$ of dipoles, with average
bending angle $\theta = 2\pi/MN_\textrm{cells}$.  If the two outer dipoles have
bending angle $a\theta$ and the inner dipoles have bending angle $b\theta$,
then the coefficients $a$ and $b$ satisfy:
\begin{equation}
2a + (M-2)b = M.
\end{equation}
Let us assume that the lattice functions (Courant--Snyder parameters and
dispersion) in the outer dipoles are the same as in a DBA lattice, and in the
inner dipoles are the same as in a TME lattice.
Since the synchrotron radiation integrals are additive, for an $M$-bend
achromat, we can write:
\begin{eqnarray}
I_{5,\textrm{cell}} & \approx &
\frac{2}{4\sqrt{15}} \frac{(a \theta)^4}{\rho} +
\frac{(M-2)}{12\sqrt{15}} \frac{(b \theta)^4}{\rho} =
\frac{6a^4 + (M-2)b^4}{12\sqrt{15}} \frac{\theta^4}{\rho}, 
 \\
I_{2,\textrm{cell}} & \approx &
2 \frac{a \theta}{\rho} + (M-2) \frac{b \theta}{\rho} =
\left( 2a + (M-2) b \right) \frac{\theta}{\rho}.
\end{eqnarray}
Hence, in an $M$-bend achromat:
\begin{equation}
\frac{I_{5,\textrm{cell}}}{I_{2,\textrm{cell}}} \approx
\frac{1}{12\sqrt{15}} \left(
\frac{6a^4 + (M-2)b^4}{2a + (M-2)b}
\right) \theta^3.
\end{equation}
Minimising the ratio $I_5/I_2$ with respect to $a$ gives:
\begin{equation}
\frac{a}{b} = \frac{1}{\sqrt[3]{3}},
\end{equation}
from which it follows that:
\begin{equation}
\left( \frac{6a^4 + (M-2)b^4}{2a + (M-2)b} \right) \approx
\frac{M+1}{M-1}.
\end{equation}
The central bending magnets should be longer than the outer bending magnets
by a factor $\sqrt[3]{3}$.
Then, the minimum natural emittance in an $M$-bend achromat is given by:
\begin{equation}
\varepsilon_{0,\textrm{MBA,min}} \approx \frac{1}{12\sqrt{15}}
\left( \frac{M+1}{M-1} \right) C_q \gamma^2 \theta^3.
\label{mbaemittance}
\end{equation}
Note that $\theta$ is the \emph{average} bending angle per dipole.
Although we derived (\ref{mbaemittance}) with the assumption of at
least three dipoles ($M>2$), the formula gives the correct result for
a DBA in the case $M = 2$.  Also, in the limit $M \to \infty$, we obtain
the correct expression for the natural emittance in a TME lattice.










Triple bend achromats have been used in light sources, including the ALS
\cite{jackson1993} and the SLS \cite{streun2001}.  Light sources based on cells with
even larger numbers of bends per achromat are planned: see, for example, \cite{leemann2009}.
As with double bend achromats, it is possible to obtain some reduction in the natural
emittance of a triple (or higher) bend achromat by detuning the lattice
from the strict achromat condition, allowing some dispersion to `leak' into the
straight sections.  As long as the dispersion in the straights is not too large, there
is a net benefit, despite some contribution to the emittance from quantum excitation
in the insertion devices.

As a final remark, we note that further flexibility to optimise the natural emittance
can be provided by relaxing the constraint that the field strength in a dipole is
constant along the length of the dipole.
We expect an optimised design to have
the strongest field at the centre of the dipole, where the dispersion can be
minimised.  For an example, see \cite{guoraubenheimer2002}.

\begin{table}[t]
\begin{center}
\caption{Minimum natural emittance in different lattice styles for electron storage rings:
for each lattice style, the minimum natural emittance is given by $\mathcal{F}C_q \gamma^2 \theta^3$,
where $C_q \approx 3.832\times 10^{-13}\,$m, and $\gamma$ is the relativistic factor for
the beam.  The dipoles have length $L$ and bending angle $\theta$, and no quadrupole component.}
\begin{tabular}{ccc}
\hline\hline
\textbf{Lattice style} & $\mathcal{F}$ & \textbf{Conditions} \\
\hline
90$^\circ$ FODO & $2\sqrt{2}$ & $f = L/\sqrt{2}$ \\
137$^\circ$ FODO & $1.2$ & minimum emittance FODO \\
DBA & $\frac{1}{4\sqrt{15}}$ & 
$\eta_{x,0} = \eta_{px,0} = 0, \quad
\beta_{x,0} \approx \sqrt{12/5} L \quad \alpha_{x,0} \approx \sqrt{15}$ \\
MBA & $\frac{1}{12\sqrt{15}} \left( \frac{M+1}{M-1} \right)$ & 
$M$ dipoles (with same radius of curvature) per cell \\
TME & $\frac{1}{12\sqrt{15}}$ &
$\eta_{x,\textrm{min}} \approx \frac{L\theta}{24} \quad
\beta_{x,\textrm{min}} \approx \frac{L}{2\sqrt{15}}$
\\
\hline\hline
\end{tabular}
\end{center}
\end{table}

\section{Vertical emittance generation, calculation and tuning\label{section3}}

In this section, we shall discuss how vertical emittance is generated by alignment
and tuning errors, describe methods for calculating the vertical emittance in the
presence of known errors, and discuss briefly how an operating storage ring can
be tuned to minimise the vertical emittance (even when the alignment and tuning
errors are not well known).

Recall that the natural (horizontal) emittance in a storage ring is given by
(\ref{naturalemittance1}):
\begin{equation}
\varepsilon_0 = C_q \gamma^2 \frac{I_5}{j_x I_2}.
\label{naturalemittance}
\end{equation}
If the horizontal and vertical motion are independent of each other (i.e.~if
there is no betatron coupling) then we can apply the same analysis to the
vertical motion as we did to the horizontal.
If we build a ring that is completely flat (i.e.~no vertical bending), then
there is no vertical dispersion, i.e.~$\eta_y = \eta_{py} = 0$ at all locations
around the ring.  It follows that the vertical $\mathcal{H}$ function $\mathcal{H}_y$:
\begin{equation}
\mathcal{H}_y = \gamma_y \eta_y^2 + 2\alpha_y \eta_y \eta_{py} + \beta_y \eta_{py}^2,
\end{equation}
also vanishes around the entire ring, and that therefore the synchrotron
radiation integral $I_{5y}$  will be zero.
This implies that the vertical emittance will damp to zero.


However, in deriving equation (\ref{naturalemittance}) for the natural emittance,
we assumed that all photons were emitted directly along the instantaneous direction
of motion of the electron.  In fact, photons are emitted with a distribution having
angular width $1/\gamma$ about the direction of motion of the electron.  This leads
to some vertical `recoil' that excites vertical betatron motion, resulting in a non-zero
vertical emittance.  A detailed analysis leads to the following formula for the
fundamental lower limit on the vertical emittance \cite{raubenheimer1991}:
\begin{equation}
\varepsilon_{y,\textrm{min}} = \frac{13}{55} \frac{C_q}{j_y I_2} \oint \frac{\beta_y}{| \rho |^3} \,ds.
\label{verticalemittancelowerlimit}
\end{equation}
To estimate a typical value for the lower limit on the vertical emittance, let us
write equation (\ref{verticalemittancelowerlimit}) in the approximate form:
\begin{equation}
\varepsilon_{y,\textrm{min}} \approx \frac{C_q \langle \beta_y \rangle}{4j_y I_2} \oint \frac{1}{| \rho |^3} \,ds
= \frac{\langle \beta_y \rangle}{4} \frac{j_z}{j_y} \frac{\sigma_\delta^2}{\gamma^2},
\end{equation}
where $\langle \beta_y \rangle$ is the average vertical beta function around
the ring.
Using some typical values ($\langle \beta_y \rangle = 20$\,m, $j_z = 2$, $j_y = 1$,
$\sigma_\delta = 10^{-3}$, $\gamma = 6000$), we find:
\begin{equation}
\varepsilon_{y,\textrm{min}} \approx 0.3\,\textrm{pm}.
\end{equation}
The lowest vertical emittance achieved so far in a storage ring is around a picometer,
several times larger than the fundamental lower limit (see, for example, \cite{dowd2011,aiba2012}).
In practice, vertical emittance in a (nominally planar) storage ring is dominated
by two effects:
residual vertical dispersion, which couples longitudinal and vertical motion;
and betatron coupling, which couples horizontal and vertical motion.
The dominant causes of residual vertical dispersion and betatron coupling are
magnet alignment errors, in particular:
tilts of the dipoles around the beam axis;
vertical alignment errors on the quadrupoles;
tilts of the quadrupoles around the beam axis;
and vertical alignment errors of the sextupoles.
Let us consider these errors in a little more detail.



Steering errors lead to a distortion of the closed orbit, which generates vertical
dispersion and (through vertical offsets of the beam in the sextupoles) betatron coupling.
A vertical steering error may be generated by
rotation of a dipole, so that the field is not exactly vertical,
or by vertical misalignment of a quadrupole, so that there is a horizontal
magnetic field at the location of the reference trajectory.


Coupling errors lead to a transfer of horizontal betatron motion and dispersion
into the vertical plane: in both cases, the result is an increase in vertical emittance.
Coupling may result from rotation of a quadrupole, so that the field contains
a skew component.
When particles pass through a skew quadrupole, they receive a vertical
kick that depends on their horizontal offset.
As a result, quantum excitation of the horizontal emittance feeds into the
vertical plane.



A vertical beam offset in a sextupole has the same effect as a skew
quadrupole.  To understand this, recall that a sextupole field is given by:
\begin{eqnarray}
B_x & = & (B\rho) k_2 xy, \\
B_y & = & \frac{1}{2} (B\rho) k_2 \left( x^2 - y^2 \right).
\end{eqnarray}
A vertical offset can be represented by the transformation $y \mapsto y + \Delta y$:
\begin{eqnarray}
B_x & \mapsto & (B\rho) k_2 xy + (B\rho) k_2 \Delta y \, x, 
\label{offsetsextupolebx} \\
B_y & \mapsto & \frac{1}{2} (B\rho) k_2 \left( x^2 - y^2 \right) - (B\rho) k_2 \Delta y \, y 
 - \frac{1}{2} k_2 \Delta y^2.
\label{offsetsextupoleby}
\end{eqnarray}
The terms in (\ref{offsetsextupolebx}) and (\ref{offsetsextupoleby}) that
are first order in $\Delta y$
constitute a skew quadrupole of strength $(B\rho) k_2 \Delta y$.

When designing and building a storage ring, we need to know how accurately
the magnets must be aligned, to keep the vertical emittance below some specified
limit.  Although beam-based tuning methods also normally have to be applied,
the ultimate emittance achieved after machine tuning does depend on the
accuracy with which the initial alignment is performed. It is therefore useful to
have expressions that relate the closed orbit distortion, vertical dispersion,
betatron coupling and (ultimately) the vertical emittance, to the alignment errors
on the magnets.

\subsection{Closed orbit distortion}

\begin{figure}[t]
\begin{center}
\includegraphics[width=0.9\textwidth]{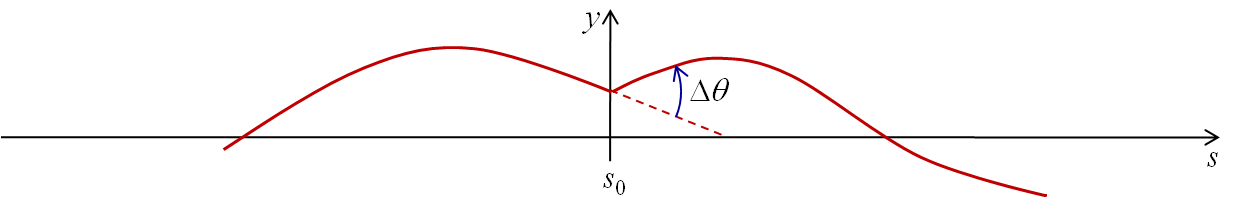}
\caption{Closed orbit distortion from a thin dipole kick in a synchrotron
storage ring.  If the co-ordinate and momentum of a particle on the closed orbit
immediately after the dipole kick are $(y_0, p_{y0})$, then after nearly
one complete turn, just before the dipole kick the co-ordinate and
momentum of the particle are $(y_0, p_{y0}- \Delta\theta)$.  The dipole kick
then puts the particle back onto the closed orbit.}
\label{closedorbitdistortion}
\end{center}
\end{figure}

Let us begin by considering the closed orbit distortion.
In terms of the action-angle variables, we can write the coordinate and momentum
of a particle at any point:
\begin{eqnarray}
y & = & \sqrt{2 \beta_y J_y} \cos \phi_y, \\
p_y & = & -\sqrt{\frac{2J_y}{\beta_y}} \left( \sin \phi_y + \alpha_y \cos \phi_y \right).
\end{eqnarray}
Suppose there is a steering error at some location $s = s_0$ which leads to an
instantaneous change (i.e.~a `kick') $\Delta \theta$ in the vertical momentum.
After one complete turn of the storage ring, starting from immediately after $s_0$,
the trajectory of a  particle will close on itself if:
\begin{eqnarray}
\sqrt{2 \beta_{y0} J_{y0}} \cos \phi_{y1} & = & \sqrt{2 \beta_{y0} J_{y0}} \cos \phi_{y0}, \label{cody} \\
-\sqrt{\frac{2J_{y0}}{\beta_{y0}}} \left( \sin \phi_{y1} + \alpha_{y0} \cos \phi_{y1} \right) & = &
-\sqrt{\frac{2J_{y0}}{\beta_{y0}}} \left( \sin \phi_{y0} + \alpha_{y0} \cos \phi_{y0} \right) - \Delta \theta. \nonumber \\
 & & \label{codpy}
\end{eqnarray}
where $\phi_{y1} = \phi_{y0} + 2\pi\nu_y$, and $\nu_y = \mu_y /2\pi$ is the vertical tune
(see Fig.~\ref{closedorbitdistortion}).
Solving equations (\ref{cody}) and (\ref{codpy}) for the action and angle at $s_0$:
\begin{eqnarray}
J_{y0} & = & \frac{\beta_{y0} \Delta \theta^2}{8\sin^2 \pi \nu_y}, \\
\phi_{y0} & = & \pi \nu_y.
\label{actionangleclosedorbit}
\end{eqnarray}
Note that if the tune is an integer, there is no solution for the closed orbit: even
the smallest steering error will kick the beam out of the ring.
From (\ref{actionangleclosedorbit}), we
can write the coordinate for the closed orbit at any point in the ring:
\begin{equation}
y_\mathrm{co}(s) = \frac{\sqrt{\beta_y(s_0)\beta_y(s)}}{2\sin \pi \nu_y} \Delta \theta
\cos \!\left( \pi \nu_y + \mu_y(s;s_0) \right),
\end{equation}
where $\mu_y(s;s_0)$ is the phase advance from $s_0$ to $s$.  

In general, there will be many steering errors distributed around a storage ring.
The closed orbit can be found by summing the effects of all the steering errors:
\begin{equation}
y_\mathrm{co}(s) = \frac{\sqrt{\beta_y(s)}}{2\sin \pi \nu_y} \oint \sqrt{\beta_y(s^\prime}) \frac{d\theta}{ds^\prime}
\cos \!\left( \pi \nu_y + \mu_y(s;s^\prime) \right) \, ds^\prime.
\label{closedorbitequation}
\end{equation}

It is often helpful to be able to estimate the size of the closed orbit distortion
that may be expected from random quadrupole misalignments of a given magnitude.
We can derive an expression for this from equation (\ref{closedorbitequation}).
For a quadrupole of integrated focusing strength $k_1 L$, vertically misaligned from
the reference trajectory by $\Delta Y$, the steering is:
\begin{equation}
\Delta \theta = (k_1 L) \Delta Y.
\end{equation}
Squaring equation (\ref{closedorbitequation}), then averaging over many seeds of random
alignment errors, we find:
\begin{equation}
\left\langle \frac{y_\mathrm{co}^2 (s)}{\beta_y (s)} \right\rangle =
\frac{\langle \Delta Y^2 \rangle}{8\sin^2 \pi \nu_y} \sum_\textrm{quads} \beta_y (k_1 L)^2.
\label{closedorbitmagnetalignment}
\end{equation}
In performing the average, we assume that the alignments of different
quadrupoles are not correlated in any way.

The ratio between the closed orbit rms and the magnet misalignment rms is sometimes
known as the \emph{orbit amplification factor}.
Values for the orbit amplification factor are typically in the range from 10 to about 100.
Of course, the amplification factor is a statistical quantity: the actual rms of the
orbit distortion depends on the particular set of alignment errors present.

\begin{figure}[t]
\begin{center}
\includegraphics[width=0.8\textwidth]{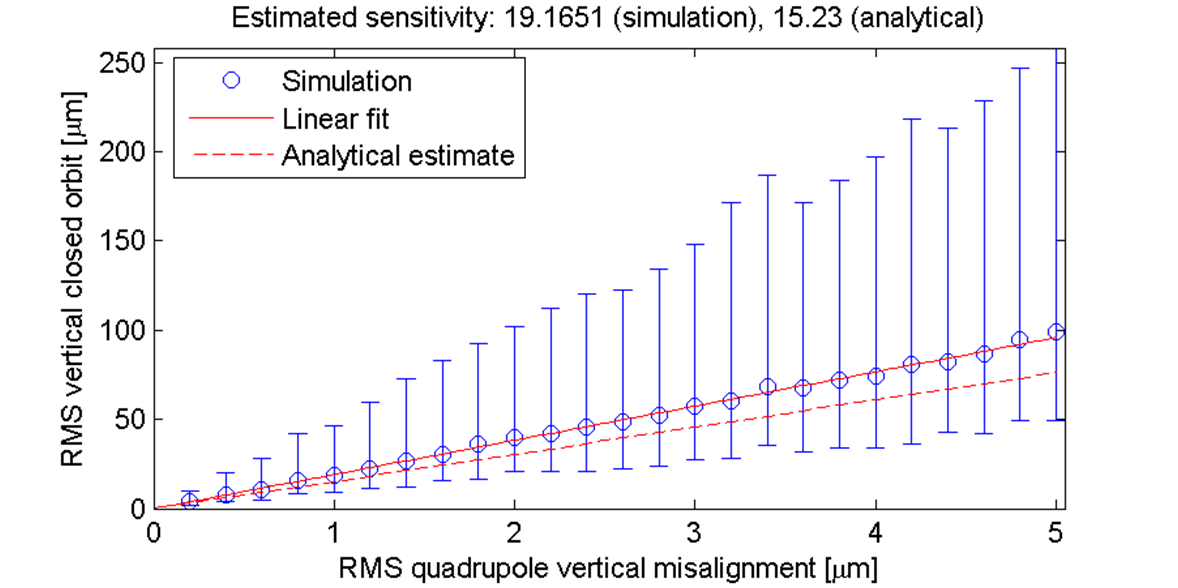}
\caption{Simulation of closed orbit distortion resulting from quadrupole
alignment errors in a storage ring \cite{wolski2006a}.  Each circle shows the mean of the
rms orbit distortion from 100 different sets (seeds) of random alignment errors
on the quadrupoles; the error bars show the range covered by 90\% of
the seeds.  The solid red line shows a linear fit to the circles; the broken
red line shows an analytical estimate of the orbit distortion based on the
known quadrupole strengths and lattice functions, using equation
(\ref{closedorbitmagnetalignment}).}
\label{figquadalignmenterror}
\end{center}
\end{figure}

In the context of low-emittance storage rings, vertical closed orbit errors are
of concern for two reasons.
First, vertical steering generates vertical dispersion, which is a source of
vertical emittance.
Second, vertical orbit errors contribute to vertical beam offset in the sextupoles,
which effectively generates skew quadrupole fields, which in turn lead to
betatron coupling.
We have seen how to analyse the beam dynamics to understand the closed orbit
distortion that arises from quadrupole alignment errors of a given magnitude.
Our goal is to relate quantities such as orbit distortion, vertical dispersion,
coupling, and vertical emittance, to the alignment errors on the magnets.
We continue with betatron coupling.

\subsection{Betatron coupling}

Betatron coupling describes the effects that can arise when the vertical motion
of a particle depends on its horizontal motion, and vice-versa.  Betatron coupling
can arise (for example) from skew quadrupoles and solenoids.


In a storage ring, skew quadrupole fields ofen arise from quadrupole tilts, and
from vertical alignment errors on sextupoles.
A full treatment of betatron coupling can become quite complex, and there
are many different formalisms that can be used.
However, it is possible to use a simplified model to derive approximate
expressions the equilibrium emittances in the presence of coupling.
The procedure is as follows.
First, we write down the equations of motion for a single particle in a beamline
containing coupling.
Then, we look for a `steady state' solution to the equations of motion, in which the horizontal
and vertical actions are each constants of the motion.
Finally, we assume that the actions in the steady state solution correspond to the equilibrium
emittances (since $\varepsilon = \langle J \rangle$), and that the
sum of the horizontal and vertical emittances is equal to the natural emittance of the `ideal'
lattice (i.e.~the natural emittance of the lattice in the absence of errors).
This procedure can give some useful results, but because of the approximations
involved, the formulae are not always very accurate.

We will use Hamiltonian mechanics.  In this formalism, the equations of motion
for the action-angle variables (with path length $s$ as the independent variable)
are derived from the Hamiltonian:
\begin{equation}
H = H(\phi_x, J_x, \phi_y, J_y; s),
\end{equation}
using Hamilton's equations:
\begin{eqnarray}
\frac{dJ_x}{ds} & = & -\frac{\partial H}{\partial \phi_x}, \\
\frac{dJ_y}{ds} & = & -\frac{\partial H}{\partial \phi_y}, \\
\frac{d\phi_x}{ds} & = & \frac{\partial H}{\partial J_x}, \\
\frac{d\phi_y}{ds} & = & \frac{\partial H}{\partial J_y}.
\end{eqnarray}
For a particle moving along a linear, uncoupled beamline, the Hamiltonian is:
\begin{equation}
H = \frac{J_x}{\beta_x} + \frac{J_y}{\beta_y}.
\end{equation}

The first step is to derive an appropriate form for the Hamiltonian in a storage
ring with skew quadrupole perturbations.
In Cartesian variables, the equations of motion in a skew quadrupole can be written:
\begin{eqnarray}
\frac{dp_x}{ds} & = & k_s y, \\
\frac{dp_y}{ds} & = & k_s x, \\
\frac{dx}{ds} & = & p_x, \\
\frac{dy}{ds} & = & p_y,
\end{eqnarray}
where:
\begin{equation}
k_s = \frac{1}{B\rho} \frac{\partial B_x}{\partial x}.
\end{equation}
These equations can be derived from the Hamiltonian:
\begin{equation}
H = \frac{1}{2}p_x^2 + \frac{1}{2}p_y^2 - k_s xy.
\end{equation}

We are interested in the case where there are skew quadrupoles distributed
around a storage ring.
The `focusing' effect of a skew quadrupole is represented by a term in the
Hamiltonian:
\begin{equation}
k_s xy = 2k_s \sqrt{\beta_x \beta_y} \sqrt{J_x J_y} \cos \phi_x \cos \phi_y.
\end{equation}
This implies that the Hamiltonian for a beam line with distributed skew quadrupoles
can be written:
\begin{equation}
H = \frac{J_x}{\beta_x} + \frac{J_y}{\beta_y} - 2k_s (s) \sqrt{\beta_x \beta_y} \sqrt{J_x J_y} \cos \phi_x \cos \phi_y.
\end{equation}
The beta functions and the skew quadrupole strength are functions of the position $s$.
This makes it difficult to solve the equations of motion exactly.
Therefore, we simplify the problem by `averaging' the Hamiltonian:
\begin{equation}
H = \omega_x J_x + \omega_y J_y - 2\bar{\kappa} \sqrt{J_x J_y} \cos \phi_x \cos \phi_y.
\end{equation}
Here, $\omega_x$, $\omega_y$ are the phase advances per unit length of the beam line, given by:
\begin{equation}
\omega_{x,y} = \frac{1}{C_0} \int_0^{C_0} \frac{ds}{\beta_{x,y}},
\end{equation}
where $C_0$ is the circumference of the ring.
$\bar{\kappa}$ is a constant that characterises the coupling strength.
For reasons that will become clear shortly, we re-write the coupling term, to
put the Hamiltonian in the form:
\begin{equation}
H = \omega_x J_x + \omega_y J_y - \bar{\kappa}_- \sqrt{J_x J_y} \cos (\phi_x - \phi_y) - \bar{\kappa}_+ \sqrt{J_x J_y} \cos (\phi_x + \phi_y).
\end{equation}
The constants $\bar{\kappa}_\pm$ represent the skew quadrupole strength averaged around
the ring.  However, we need to take into account that the kick from a skew quadrupole
depends on the betatron phase.  Thus, we write:
\begin{equation}
\bar{\kappa}_\pm e^{i\chi} = \frac{1}{C_0} \int_0^{C_0} e^{i(\mu_x \pm \mu_y)} k_s \sqrt{\beta_x \beta_y}\, ds,
\label{couplingstrength1}
\end{equation}
where $\mu_x$ and $\mu_y$ are the betatron phase advances from the start of the ring.

Now suppose that $\bar{\kappa}_- \gg \bar{\kappa}_+$.  (This can occur, for example,
if $\omega_x \approx \omega_y$, in which case all the contributions to $\bar{\kappa}_-$
from the skew quadrupole perturbations will add together in phase.)  Then, we can simplify
things further by dropping the term in $\bar{\kappa}_+$ from the Hamiltonian:
\begin{equation}
H = \omega_x J_x + \omega_y J_y - \bar{\kappa}_- \sqrt{J_x J_y} \cos (\phi_x - \phi_y).
\label{couplinghamiltonian}
\end{equation}
We can now write down the equations of motion:
\begin{eqnarray}
\frac{dJ_x}{ds} & = & -\frac{\partial H}{\partial \phi_x} = \bar{\kappa}_- \sqrt{J_x J_y} \sin (\phi_x - \phi_y), \label{coupledeom1} \\
\frac{dJ_y}{ds} & = & -\frac{\partial H}{\partial \phi_y} = -\bar{\kappa}_- \sqrt{J_x J_y} \sin (\phi_x - \phi_y), \label{coupledeom2} \\
\frac{d\phi_x}{ds} & = & \frac{\partial H}{\partial J_x} = \omega_x + \frac{\bar{\kappa}_-}{2} \sqrt{\frac{J_x}{J_y}} \cos (\phi_x - \phi_y), \label{coupledeom3} \\
\frac{d\phi_y}{ds} & = & \frac{\partial H}{\partial J_y} = \omega_y + \frac{\bar{\kappa}_-}{2} \sqrt{\frac{J_y}{J_x}} \cos (\phi_x - \phi_y). \label{coupledeom4} 
\end{eqnarray}
 
Even after all the simplifications we have made, the equations of motion are still
rather difficult to solve.  Fortunately, however, we do not require the general solution.
In fact, we are only interested in the properties of some special cases.
First of all, we note that from (\ref{coupledeom1}) and (\ref{coupledeom2}):
\begin{equation}
\frac{dJ_x}{ds} + \frac{dJ_y}{ds} = 0,
\end{equation}
and therefore the sum of the actions $J_x + J_y$ is constant.
Going further, we notice that if $\phi_x = \phi_y$, then the rate of change of each action
falls to zero.
This implies that if we can find a solution to the equations of motion with
$\phi_x = \phi_y$ for all $s$, then the actions will remain constant.
In fact, we find that if $\phi_x = \phi_y$, and:
\begin{equation}
\frac{d\phi_x}{ds} = \frac{d\phi_y}{ds},
\end{equation}
then:
\begin{equation}
\frac{J_y}{J_x} = \frac{\sqrt{1 + \bar{\kappa}_-^2 / \Delta \omega^2} - 1}{\sqrt{1 + \bar{\kappa}_-^2 / \Delta \omega^2} + 1},
\end{equation}
where $\Delta \omega = \omega_x - \omega_y$.
If we further use $J_x + J_y = J_0$, where $J_0$ is a constant, then we have a
solution to the equations of motion in which the actions are constant, and given by:
\begin{eqnarray}
J_x & = & \frac{1}{2} \left( 1 + \frac{1}{\sqrt{1 + \bar{\kappa}_-^2 / \Delta \omega^2}} \right) J_0,
\label{fixedpointactionx}\\
J_y & = & \frac{1}{2} \left( 1 - \frac{1}{\sqrt{1 + \bar{\kappa}_-^2 / \Delta \omega^2}} \right) J_0.
\label{fixedpointactiony}
\end{eqnarray}

\begin{figure}[t]
\begin{center}
\includegraphics[width=0.5\textwidth]{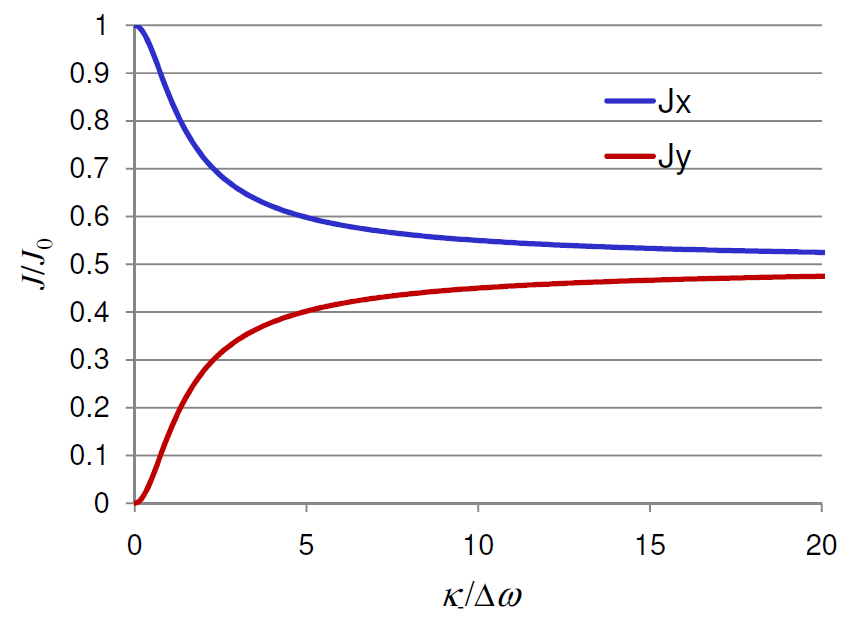}
\caption{Variation of the `fixed point' actions (\ref{fixedpointactionx}) and
(\ref{fixedpointactiony}) as a function of the strength of the coupling resonance.}
\label{figcouplingresonance}
\end{center}
\end{figure}

Note the behaviour, shown in Fig.~\ref{figcouplingresonance}, of the fixed
actions as we vary the `coupling strength' $\bar{\kappa}_-$ and the betatron
tunes (betatron frequencies). The fixed actions are well-separated for
$\bar{\kappa}_- \ll \Delta \omega$, but both approach the value $J_0/2$ for
$\bar{\kappa}_- \gg \Delta \omega$.  The condition at which the tunes are
equal (or differ by an exact integer) is known as the
\emph{difference coupling resonance}.

Recall that the emittance may be defined as the betatron action averaged
over all particles in the beam:
\begin{equation}
\varepsilon_x = \langle J_x \rangle, \qquad \textrm{and} \qquad
\varepsilon_y = \langle J_y \rangle.
\end{equation}
Now, synchrotron radiation will damp the beam towards an equilibrium
distribution.  In this equilibrium, we expect the betatron actions of the particles
to change only slowly, i.e.~on the timescale of the radiation damping, which is
much longer than the timescale of the betatron motion.
In that case, the actions of most particles must be in the correct ratio for a 
fixed-point solution to the equations of motion.  Then, if we assume that
$\varepsilon_x + \varepsilon_y = \varepsilon_0$, where $\varepsilon_0$ is
the natural emittance of the storage ring, we must have for the equilibrium
emittances:
\begin{eqnarray}
\varepsilon_x & = & \left( 1 + \frac{1}{\sqrt{1 + \bar{\kappa}_-^2 / \Delta \omega^2}} \right) \frac{\varepsilon_0}{2} ,
\label{verticalemittancebetatroncoupling1} \\
\varepsilon_y & = & \left( 1 - \frac{1}{\sqrt{1 + \bar{\kappa}_-^2 / \Delta \omega^2}} \right) \frac{ \varepsilon_0}{2} .
\label{verticalemittancebetatroncoupling2}
\end{eqnarray}

\begin{figure}[t]
\begin{center}
\includegraphics[width=0.6\textwidth]{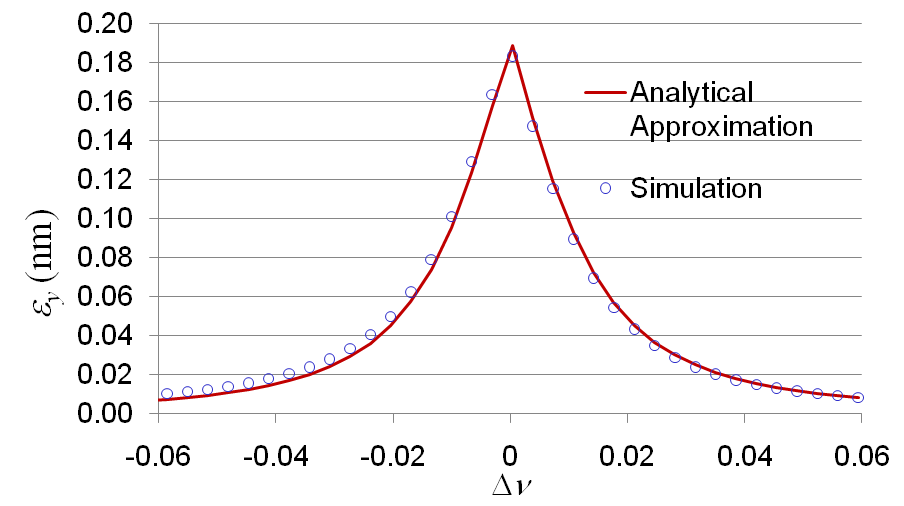}
\caption{Effect of a single skew quadrupole (at a location with zero dispersion)
on the vertical emittance in a synchrotron storage ring, as a function of the difference
in the betatron tunes.  The circles show the results of a computation using Chao's
method \cite{chao1979}; the red line shows an analytical estimate using equation
(\ref{verticalemittancebetatroncoupling2}).}
\label{figcouplingresonance1}
\end{center}
\end{figure}

As an illustration, we can plot the vertical emittance as a function of the
`tune split' $\Delta \nu$, in a model of the ILC damping rings, with a single
skew quadrupole (located at a point of zero dispersion, so as not to couple horizontal
dispersion into the vertical plane).  The result is shown in Fig.~\ref{figcouplingresonance1}.
The tunes are controlled by adjusting the regular (normal) quadrupoles in the lattice.
The simulation results are based on emittance calculation using
Chao's method, which we shall discuss later.

The presence of skew quadrupole errors in a storage ring affects the betatron
tunes.  To estimate the size of the effect,
we use the Hamiltonian (\ref{couplinghamiltonian}).  If we consider a particle close
to the fixed point solution, we can assume that $\phi_x = \phi_y$, so that
the Hamiltonian becomes:
\begin{equation}
H = \omega_x J_x + \omega_y J_y - \bar{\kappa}_- \sqrt{J_x J_y}.
\label{couplinghamiltonian2}
\end{equation}
The normal modes describe motion that is periodic with a single well-defined frequency.
In the absence of coupling, the transverse normal modes correspond to motion in
just the horizontal or vertical plane.  When coupling is present, the normal modes
involve a combination of horizontal and vertical motion.

\begin{figure}[t]
\begin{center}
\includegraphics[width=0.6\textwidth]{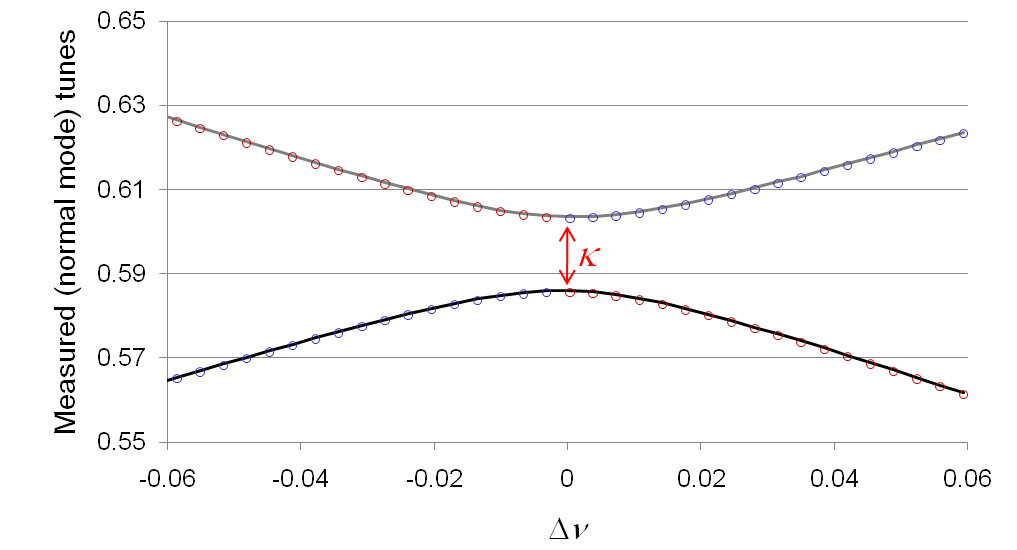}
\caption{Effect of a single skew quadrupole (at a location with zero dispersion)
on the measured betatron tunes in a synchrotron storage ring, as a function of the
difference in the betatron tunes in the absence of the skew quadrupole.  The circles
show the results of a computation of the eigenvalues of the single-turn transfer
matrix; the solid lines show an analytical estimate using equation
(\ref{betatrontunescouplingerror}).  The minimum difference between the
measured tunes gives the coupling strength in the ring.}
\label{figcouplingresonance2}
\end{center}
\end{figure}

Let us write the Hamiltonian (\ref{couplinghamiltonian2}) in the form:
\begin{equation}
H = \left( \begin{array}{cc} \sqrt{J_x} & \sqrt{J_y} \end{array} \right)
A
\left( \begin{array}{c} \sqrt{J_x} \\ \sqrt{J_y} \end{array} \right),
\end{equation}
where:
\begin{equation}
A = \left( \begin{array}{cc} \omega_x & -\frac{1}{2}\bar{\kappa}_- \\ 
                                   -\frac{1}{2}\bar{\kappa}_- & \omega_y \end{array} \right).
\end{equation}
The normal modes can be constructed from the eigenvectors of the matrix $A$,
and the frequency of each mode is given by the corresponding eigenvalue.
From the eigenvalues of $A$, we find that the normal mode frequencies are:
\begin{equation}
\omega_\pm = \frac{1}{2} \left(
\omega_x + \omega_y \pm \sqrt{\bar{\kappa}_-^2 + \Delta \omega^2}
\right).
\end{equation}
Hence, the tunes $\nu_\pm$ are given (in terms of the tunes $\nu_x$ and $\nu_y$
in the absence of errors) by:
\begin{equation}
\nu_\pm = \frac{1}{2} \left( \nu_x + \nu_y \pm \sqrt{\kappa^2 + \Delta \nu^2} \right),
\label{betatrontunescouplingerror}
\end{equation}
where, from (\ref{couplingstrength1}), $\kappa = (C_0/2\pi) \bar{\kappa}_-$ is given by:
\begin{equation}
\kappa e^{i\chi} = \frac{1}{2\pi} \int_0^{C_0} e^{i(\mu_x - \mu_y)} k_s \sqrt{\beta_x \beta_y}\, ds.
\label{couplingstrength2}
\end{equation}
The dependence of the tunes on the coupling strength provides a useful method
for measuring the coupling strength $\kappa$ in a real lattice.  The procedure is
simple: a quadrupole (or combination of quadrupoles) is used to change the
tunes, and then the tunes are recorded and plotted as a function of quadrupole
strength.  The minimum separation between the measured tunes gives the
coupling strength.  An example (from simulation) is shown in Fig.~\ref{figcouplingresonance2}.
Of course, this procedure does not identify the source of the
coupling, or provide very much information as to an optimal correction (beyond the
strength of a skew quadrupole that may be required to achieve the correction,
assuming that the skew quadrupole is at the correct phase in the lattice).  However,
the technique can be useful to characterise the effect of a correction that may
need to be applied in several iterations.

Major sources of coupling in storage rings are quadrupole tilts and sextupole
alignment.  Using the theory just outlined, we can estimate the alignment
tolerances on these magnets, for given optics and specified vertical emittance.
Starting with equation (\ref{couplingstrength2}), we first take the modulus squared,
and then use (for a sextupole with vertical alignment error $\Delta Y_S$)
$k_s = k_2 \Delta Y_S$ and (for a quadrupole with tilt error $\Delta \Theta_Q$)
$k_s = k_1 \Delta\Theta_Q$.  Assuming that there are no correlations between
the errors, we find:
\begin{equation}
\langle \kappa^2 \rangle \approx
\frac{\langle \Delta Y_S^2 \rangle}{4\pi^2}
\sum_\mathrm{sexts} \beta_x \beta_y (k_2 l)^2 +
\frac{\langle \Delta \Theta_Q^2 \rangle}{4\pi^2}
\sum_\mathrm{quads} \beta_x \beta_y (k_1 l)^2,
\label{betatroncouplingalignmenterrors}
\end{equation}
where $\langle \kappa^2 \rangle$ represents the mean value of the square of
the coupling strength over a large number of sets of random errors.
Note that $\Delta Y_S$ is the beam offset from the centre of a sextupole:
this includes the effects of closed orbit distortion.

\subsection{Vertical dispersion}

Vertical emittance is generated by vertical dispersion as well as by betatron
coupling.
Vertical dispersion is in turn generated by
vertical closed orbit distortion (vertical steering), and
coupling of horizontal dispersion into the vertical plane by skew quadrupole
fields.
Our goal now is to estimate the amount of vertical dispersion generated from magnet alignment
errors; we can then estimate the contribution to the vertical emittance.

The equation of motion for the vertical co-ordinate for a particle with momentum $P$ is:
\begin{equation}
\frac{d^2 y}{ds^2} = \frac{B_x}{(B\rho)} = \frac{q}{P} B_x.
\end{equation}
For small energy deviation $\delta$, $P$ is related to the reference momentum $P_0$ by:
\begin{equation}
P \approx (1 + \delta) P_0.
\end{equation}
We can write for the horizontal field (to first order in the derivatives):
\begin{equation}
B_x \approx B_{0x} + y \frac{\partial B_x}{\partial y} + x \frac{\partial B_x}{\partial x}.
\end{equation}
If we consider a particle following an off-momentum closed orbit, so that:
\begin{eqnarray}
y & = & \eta_y \delta, \\
x & = & \eta_x \delta,
\end{eqnarray}
then, combining the above equations, we find to first order in $\delta$:
\begin{equation}
\frac{d^2 \eta_y}{ds^2} - k_1 \eta_y \approx -k_{0s} + k_{1s} \eta_x.
\label{dispersionequationofmotion}
\end{equation}

Equation (\ref{dispersionequationofmotion}) gives the `equation of motion'
for the dispersion.  It is similar to the equation of motion for the closed orbit:
\begin{equation}
\frac{d^2 y_\mathrm{co}}{ds^2} - k_1 y_\mathrm{co} \approx -k_{0s} + k_{1s} x_\mathrm{co}.
\end{equation}
We can therefore immediately generalise the relationship (\ref{closedorbitmagnetalignment})
between the closed
orbit and the quadrupole misalignments, to find for the dispersion:
\begin{equation}
\left\langle \frac{\eta_y^2}{\beta_y} \right\rangle =
\frac{\langle \Delta Y_Q^2 \rangle}{8\sin^2 \pi \nu_y} \sum_\textrm{quads} \beta_y (k_1 L)^2 +
\frac{\langle \Delta \Theta_Q^2 \rangle}{8\sin^2 \pi \nu_y} \sum_\textrm{quads} \eta_x^2 \beta_y (k_1 L)^2 +
\frac{\langle \Delta Y_S^2 \rangle}{8\sin^2 \pi \nu_y} \sum_\textrm{sexts} \eta_x^2 \beta_y (k_2 L)^2.
\label{dispersionmagnetalignment}
\end{equation}
Here, we assume that the skew dipole terms $k_{0s}$ come from vertical alignment errors
on the quadrupoles with mean square $\langle \Delta Y_Q^2 \rangle$, and that the skew
quadrupoles $k_{1s}$ come from tilts on the quadrupoles with mean square
$\langle \Delta \Theta_Q^2 \rangle$ and from vertical alignment errors on the sextupoles, with
mean square $\langle \Delta Y_S^2 \rangle$.  We assume that all alignment errors are
uncorrelated.

The final step is to relate the vertical dispersion to the vertical emittance.
This is not too difficult.
First, we can apply the formula (\ref{naturalemittance1}) for the natural (horizontal) emittance
to the vertical emittance:
\begin{equation}
\varepsilon_y = C_q \gamma^2 \frac{I_{5y}}{j_y I_2},
\end{equation}
where $j_y$ is the vertical damping partition number (usually, $j_y =1$), and the
synchrotron radiation integrals are given by:
\begin{equation}
I_{5y} = \oint \frac{\mathcal{H}_y}{| \rho |^3}\, ds,
\end{equation}
and:
\begin{equation}
I_2 = \oint \frac{1}{\rho^2} \, ds.
\end{equation}
The vertical $\mathcal{H}$ function is:
\begin{equation}
\mathcal{H}_y = \gamma_y \eta_y^2 + 2\alpha_y \eta_y \eta_{py} + \beta_y \eta_{py}^2.
\end{equation}

If the vertical dispersion is generated randomly, then we can assume that it will
\emph{not} be correlated with the curvature $1/\rho$ of the reference
trajectory.  (This is not the case for the horizontal dispersion!)
Then, we can write:
\begin{equation}
I_{5y} \approx \langle \mathcal{H}_y \rangle \oint \frac{1}{| \rho |^3} \, ds
= \langle \mathcal{H}_y \rangle I_3.
\end{equation}
Hence, for the vertical emittance:
\begin{equation}
\varepsilon_y \approx C_q \gamma^2 \langle \mathcal{H}_y \rangle \frac{I_3}{j_y I_2}.
\end{equation}
It is convenient to use (\ref{naturalenergyspreadmeansquare}) for the mean square
energy spread, to give:
\begin{equation}
\varepsilon_y \approx \frac{j_z}{j_y} \langle \mathcal{H}_y \rangle \sigma_\delta^2.
\label{verticalemittancedispersion1}
\end{equation}
Now, note the similarity between the action:
\begin{equation}
2J_y = \gamma_y y^2 + 2\alpha_y y p_y + \beta_y p_y^2,
\end{equation}
and the $\mathcal{H}$ function:
\begin{equation}
\mathcal{H}_y = \gamma_y \eta_y^2 + 2\alpha_y \eta_y \eta_{py} + \beta_y \eta_{py}^2.
\end{equation}
This implies that we can write:
\begin{equation}
\eta_y = \sqrt{\beta_y \mathcal{H}_y} \cos \phi_{\eta y},
\end{equation}
and hence:
\begin{equation}
\left\langle \frac{\eta_y^2}{\beta_y} \right\rangle = \frac{1}{2} \langle \mathcal{H}_y \rangle.
\label{verticalemittancedispersion2}
\end{equation}
Combining equations (\ref{verticalemittancedispersion1}) and (\ref{verticalemittancedispersion2})
gives a useful (approximate) relationship, between the vertical dispersion and the vertical emittance:
\begin{equation}
\varepsilon_y \approx 2 \frac{j_z}{j_y} \left\langle \frac{\eta_y^2}{\beta_y} \right\rangle \sigma_\delta^2.
\label{verticalemittancedispersion3}
\end{equation}

Equation (\ref{dispersionmagnetalignment}) tells us how the vertical dispersion
depends on the magnet alignment, and
equation (\ref{verticalemittancedispersion3}) tells us how the vertical emittance
depends on the vertical dispersion.
Simply combining these two equations gives us an equation for the contribution
of the vertical dispersion to the emittance, in terms of the magnet alignment errors.

\begin{figure}[t]
\begin{center}
\includegraphics[width=0.8\textwidth]{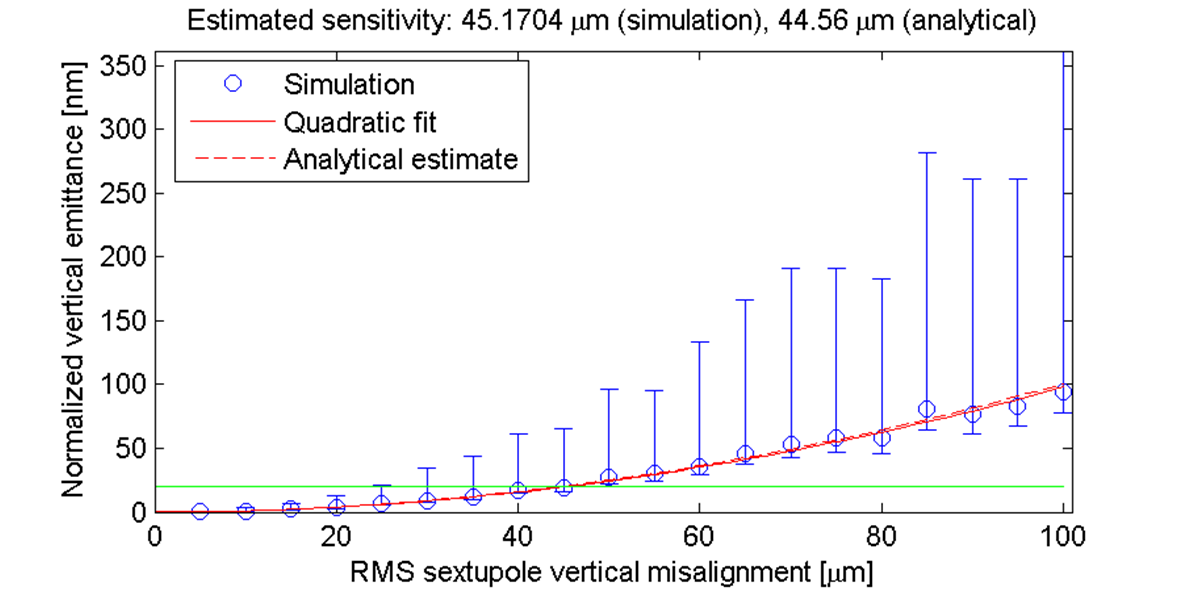}
\caption{Simulation of vertical emittance resulting from sextupole
alignment errors in a storage ring \cite{wolski2006a}.  Each circle shows the mean of the
normalised vertical emittance $(\gamma \varepsilon_y)$ from 100 different
sets (seeds) of random alignment errors on the sextupoles; the error bars show
the range covered by 90\% of the seeds.  The solid red line shows a quadratic
fit to the circles; the broken red line (very close to the solid red line) shows an
analytical estimate of the emittance based on the 
known sextupole strengths and lattice functions, using equations
(\ref{verticalemittancebetatroncoupling2}),
(\ref{betatroncouplingalignmenterrors}) to estimate the coupling contribution,
and (\ref{dispersionmagnetalignment}) and (\ref{verticalemittancedispersion3})
to estimate the dispersion contribution.}
\label{figsextalignmenterror}
\end{center}
\end{figure}

It should be remembered that the total vertical emittance is found by adding
together the contributions from betatron coupling
(equations (\ref{verticalemittancebetatroncoupling2}) and (\ref{betatroncouplingalignmenterrors}))
and vertical dispersion (equations (\ref{dispersionmagnetalignment}) and (\ref{verticalemittancedispersion3})).
All these expressions involve significant approximations.  However, they can give results
that agree reasonably well with more reliable methods: an example is shown in
Fig.~\ref{figsextalignmenterror}.

\subsection{Accurate computation of emittance}

The formulae we have derived so far are useful for developing a `feel' for how
the vertical emittance depends on magnet alignment errors, and for making rough
estimates of the sensitivity to particular types of error.
For detailed studies, including modelling and simulations, we need more accurate
formulae for computing the vertical emittance in a storage ring with a given set of
alignment errors.
The calculations involved then become quite complex, and need
to be solved using a computer.

There are three methods commonly used for computing the equilibrium emittances
in complex lattices with known errors.  First, there is a technique based on
the usual formulae for the emittances expressed in terms of the radiation integrals,
but generalised to the normal modes (see, for example, \cite{sagan2013}).  Second,
there is Chao's method \cite{chao1979}, which involves integrating the eigenvectors
of the single-turn transfer matrix around the circumference of the ring.
Finally, there is the `envelope' method \cite{ohmihirataoide1994}, in which the
second order moments of the equilibrium beam distribution are first computed from the
single-turn transfer map (including radiation damping and quantum excitation);
then the emittances are obtained from the matrix describing the beam distribution.
We shall discuss briefly each of these techniques in turn.


First, we consider the method for computing the equilibrium emittances based on
normal mode analysis.
Let us assume that we have a lattice code that will compute the \emph{symplectic}
single-turn transfer matrix at any point in a given lattice.
In general, the transfer
matrix will have non-zero terms off the block-diagonals: these terms represent
coupling between the horizontal, vertical, and longitudinal motion.
The expression (\ref{naturalemittance}) we derived for the natural emittance assumed no betatron coupling,
and that the coupling between the horizontal and longitudinal motion was
relatively weak.
However, we can generalise the formula to the case that betatron coupling is
present.
We still need to assume that the longitudinal motion is weakly
coupled to each of the transverse degrees of freedom (i.e.~the horizontal and
vertical motion).
In that case, we can consider separately the $4\times 4$
single-turn transfer matrix $R_\perp$ describing the transverse motion:
\begin{equation}
R = \left(
\begin{array}{cc}
R_\perp & \bullet \\
\bullet     & R_\parallel
\end{array}
\right).
\end{equation}
$R_\parallel$ is a $2\times 2$ matrix describing the longitudinal motion, and we
assume we can neglect the terms represented by the bullets ($\bullet$).

Now we look for a transformation, represented by a $4\times 4$ matrix $V$,
that puts $R_\perp$ into block-diagonal form, i.e.~that `decouples' the transverse motion:
\begin{equation}
\tilde{R}_\perp = V \, R_\perp \, V^{-1} =
\left(
\begin{array}{cc}
R_\textrm{I} & 0 \\
0     & R_\textrm{II}
\end{array}
\right).
\end{equation}
$R_\textrm{I}$ and $R_\textrm{II}$ are $2\times 2$ matrices describing
betatron motion in a coordinate system in which the motion appears uncoupled.
There are various recipes for constructing the decoupling transformation $V$
(which is not unique): see, for example \cite{edwardsteng1973,saganrubin1999}.
Having obtained the matrices describing the uncoupled motion, we can derive
the Courant--Snyder parameters for the normal mode motion in the usual way.
For example, we can write:
\begin{equation}
R_\textrm{II} = \left(
\begin{array}{cc}
\cos \mu_\textrm{II} + \alpha_\textrm{II} \sin \mu_\textrm{II} & \beta_\textrm{II} \sin \mu_\textrm{II} \\
-\gamma_\textrm{II} \sin \mu_\textrm{II}  & \cos \mu_\textrm{II} - \alpha_\textrm{II} \sin \mu_\textrm{II}
\end{array}
\right),
\end{equation}
and similarly for mode I.
We can also obtain the normal mode dispersion functions, by applying the transformation
$V$ to a vector constructed from the dispersion functions in the original Cartesian co-ordinates.
Then, we can construct the $\mathcal{H}$ function for each mode; for example:
\begin{equation}
\mathcal{H}_\textrm{II} = \gamma_\textrm{II} \eta_\textrm{II}^2 +
                                     2\alpha_\textrm{II} \eta_\textrm{II} \eta_{p,\textrm{II}} +
                                       \beta_\textrm{II} \eta_{p,\textrm{II}}^2.
\end{equation}
Finally, we can write for the \emph{mode II emittance}:
\begin{equation}
\varepsilon_\textrm{II} = C_q \gamma^2 \frac{I_{5,\textrm{II}}}{I_2 - I_{4,\textrm{II}}},
\label{modeIIemittance}
\end{equation}
and similarly for mode I.

For many storage rings, equation (\ref{modeIIemittance}) works well, and gives
an accurate result.
However, if there is strong coupling between the transverse and
the longitudinal motion (which can happen, for example, for large values of the synchrotron
tune), then the approximations needed to derive equation (\ref{modeIIemittance})
start to break down.

As an alternative to the normal mode analysis, we can consider Chao's method \cite{chao1979} for
computing the emittances, which
provides a formula that can be expressed in a convenient form,
though it is not always easy to apply.  It is again based on
the single-turn transfer matrix, but it is more accurate than the `decoupling' method,
since it uses the full $6\times 6$ transfer matrix, and does not assume weak coupling
between the longitudinal and transverse motion.
We do not explain here the physics behind the formula, but simply quote the result:
\begin{equation}
\varepsilon_k = C_L \frac{\gamma^5}{c \alpha_k} \oint \frac{| E_{k\, 5}(s) |^2}{| \rho(s) |^3} \, ds,
\end{equation}
where $k = $\ I, II, III is an index that specifies a particular degree of freedom, the eigenvalues of the single-turn matrix
\emph{including radiation damping} are $e^{-\alpha_k \pm 2\pi i \nu_k}$, $E_{k\, 5}$ is
the fifth component of the $k^\textrm{th}$ eigenvector of the \emph{symplectic} single-turn matrix, and:
\begin{equation}
C_L = \frac{55}{48\sqrt{3}} \frac{r_c \hbar}{m},
\end{equation}
where $r_c$ is the classical radius and $m$ the mass of the particles in the beam.

Finally, we mention the envelope method \cite{ohmihirataoide1994}.  Like Chao's method, it gives accurate
results for the emittances even if there is strong coupling between all three degrees of freedom.
The envelope method is based on finding the equilibrium beam distribution
described by the \emph{Sigma matrix}:
\begin{equation}
\Sigma =
\left(
\begin{array}{cccccc}
\langle x^2 \rangle      & \langle x p_x \rangle        & \langle x y \rangle & \langle x p_y \rangle & \langle x z \rangle & \langle x \delta \rangle \\
\langle p_x x \rangle    & \langle p_x^2 \rangle      & \langle p_x y \rangle & \langle p_x p_y \rangle & \langle p_x z \rangle & \langle p_x \delta \rangle \\
\langle y x \rangle        & \langle y p_x \rangle       & \langle y^2 \rangle & \langle y p_y \rangle & \langle y z \rangle & \langle y \delta \rangle \\
\langle p_y x \rangle    & \langle p_y p_x \rangle    & \langle p_y y \rangle & \langle p_y^2 \rangle & \langle p_y z \rangle & \langle p_y \delta \rangle \\
\langle z x \rangle        & \langle z p_x \rangle       & \langle z y \rangle & \langle z p_y \rangle & \langle z^2 \rangle & \langle z \delta \rangle \\
\langle \delta x \rangle & \langle \delta p_x \rangle & \langle \delta y \rangle & \langle \delta p_y \rangle & \langle \delta z \rangle & \langle \delta^2 \rangle
\end{array} \right)
\end{equation}
This is a symmetric matrix, constructed from the second order moments
of all possible combinations of the dynamical variables.  For simplicity, we assume in what
follows that the first order moments are all zero, i.e.~that the closed orbit lies along the
reference trajectory.  However, the method is easily generalised to include cases where
there is closed orbit distortion.
In the absence of coupling, the Sigma matrix will be block diagonal.  We are interested in the
more general case, where coupling is present.

Under a single turn around an accelerator, $\Sigma$ transforms as:
\begin{equation}
\Sigma \mapsto R \Sigma R^\textrm{T} + D,
\end{equation}
where $R$ is the single-turn transfer matrix (including radiation damping) and
$D$ is a constant matrix representing the effects of quantum excitation.
From knowledge of the properties of synchrotron radiation, we can compute
the matrices $R$ and $D$ for a given lattice design: this will be discussed further
below, where we shall give explicit expressions for the transfer matrices in a dipole,
including radiation effects.

The \emph{equilibrium} distribution $\Sigma_\textrm{eq}$ has the property:
\begin{equation}
\Sigma_\textrm{eq} = R \Sigma_\textrm{eq} R^\textrm{T} + D.
\label{equilibriumdistributioncondition}
\end{equation}
For given $R$ and $D$, we can solve equation (\ref{equilibriumdistributioncondition})
to find $\Sigma_\textrm{eq}$,
and then from $\Sigma_\textrm{eq}$ we can find the \emph{invariant} emittances,
i.e.~the conserved quantities under symplectic transport.
For any beam distribution $\Sigma$, the invariant emittances $\varepsilon_k$ are given by:
\begin{equation}
\textrm{eigenvalues}(\Sigma S) = \pm i \varepsilon_k,
\end{equation}
where $S$ is the antisymmetric block-diagonal matrix (\ref{antisymmetricmatrixs}).
To see that this is the case, consider the (simpler) case of motion in one degree of freedom.
The Sigma matrix in this case is:
\begin{equation}
\Sigma =
\left(
\begin{array}{cc}
\langle x^2 \rangle      & \langle x p_x \rangle     \\
\langle p_x x \rangle    & \langle p_x^2 \rangle      
\end{array} \right) = 
\left(
\begin{array}{cc}
\beta_x      & -\alpha_x     \\
-\alpha_x    & \gamma_x      
\end{array} \right) \varepsilon_x.
\end{equation}
In one degree of freedom, the matrix corresponding to (\ref{antisymmetricmatrixs}) is:
\begin{equation}
S =
\left(
\begin{array}{cc}
0      & 1    \\
-1    & 0
\end{array} \right).
\end{equation}
Then, the eigenvalues of $\Sigma S$ are $\pm i \varepsilon_x$.
Now, we can show that (under certain assumptions) the emittance is conserved
as a bunch is transported along a beam line.  In any number of degrees of freedom,
the linear transformation in phase space co-ordinates of a particle in the bunch
between two points in the beam line can be represented by a matrix $R$:
\begin{equation}
\vec{x} \mapsto R \vec{x},
\label{transfermatrix1}
\end{equation}
where $\vec{x}$ is a vector whose components are the phase space variables $x_i$.

Now consider how the Sigma matrix transforms. The Sigma matrix can be written as the
product of the phase-space co-ordinates averaged over the bunch:
\begin{equation}
\Sigma_{ij} = \langle x_i x_j \rangle,
\end{equation}
where $\Sigma_{ij}$ is the $(i,j)$ component of the Sigma matrix, and the $x_i$
are the dynamical variables.  The brackets $\langle \cdot \rangle$
indicate an average over all particles in the bunch.
Then, using (\ref{transfermatrix1}), it follows that under a transformation $R$ of the
dynamical variables, the Sigma matrix transforms as:
\begin{equation}
\Sigma \mapsto R \Sigma R\transpose.
\end{equation}
Since $S$ is a constant matrix, it immediately follows that:
\begin{equation}
\Sigma S \mapsto R \Sigma R\transpose S.
\end{equation}
Then, using the fact that $R$ is symplectic (\ref{symplecticcondition}), we have:
\begin{equation}
\Sigma S \mapsto R \Sigma S R^{-1}.
\end{equation}
This is a similarity transformation of $\Sigma S$: the eigenvalues
of any matrix are conserved under a similarity transformation.  Therefore,
since the eigenvalues of $\Sigma S$ give the emittance of the bunch,
it follows that the emittances are conserved under linear, symplectic transport.

This argument applies for any number of degrees of freedom.
We define the matrix $S$ in three degrees of freedom by (\ref{antisymmetricmatrixs}).
The six eigenvalues of $\Sigma S$ are then $\pm i \varepsilon_k$, where
$k$ is an index ranging over the different degrees of freedom.
The quantities $\varepsilon_k$ are all conserved under linear, symplectic transport.
Even if, as is generally the case, the Sigma matrix is not block-diagonal
(i.e.~if there is coupling present), then we can still find three invariant
emittances using this method, without any modification.

Neglecting radiaton, if $R$ is a (symplectic) matrix that represents the linear single-turn
transformation at some point in a storage ring, then an invariant or `matched' distribution
is one that satisfies:
\begin{equation}
\Sigma \mapsto R \Sigma R\transpose = \Sigma.
\end{equation}
In general, all the particles in the bunch change position in phase space after one turn
around the ring: but for a matched distribution, the second order moments remain the same.
Although this condition determines the lattice functions (which can be found from the
\emph{eigenvectors} of $\Sigma S$), it is not sufficient to determine the emittances.
In other words, the matched distribution condition determines the \emph{shape} of the bunch,
but not the \emph{size} of the bunch.  This makes sense: after all, in a proton storage ring,
we can have a matched bunch of any emittance.
However, in an electron storage ring, we know that radiation effects will damp the emittances
to some equilibrium values.
We shall now show how to apply the concept of a matched distribution, when radiation
effects are included, to find the equilibrium emittances in an electron storage ring.

To account for radiation effects in an electron storage ring, we must make two
modifications to the single-turn transformation.
First, the matrix $R$ will no longer be symplectic: this accounts for radiation damping.
Second, as well as first order terms in the transformation (represented by the matrix $R$),
there will be zeroth order terms: these will correspond to the quantum excitation.
The condition for a matched distribution should then be written:
\begin{equation}
\Sigma = R \Sigma R\transpose + D,
\label{singleturndampingexcitation}
\end{equation}
where $R$ and $D$ are constant, non-symplectic matrices that represent the first order
and zeroth order terms in the single-turn transformation, respectively.
Equation (\ref{singleturndampingexcitation}) is sufficient to determine the Sigma matrix
uniquely -- in other words,
using just this equation (with known $R$ and $D$) we can find the bunch emittances
and the matched lattice functions.

The envelope method for finding the equilibrium emittances in a storage ring
then consists of three steps.  First, we need to 
find the first order terms $R$ and zeroth order terms $D$ in the single-turn
transformation:
\begin{equation}
\Sigma \mapsto R \Sigma R\transpose + D.
\end{equation} 
In the second step, we use the matching condition (\ref{singleturndampingexcitation})
to determine the Sigma matrix.
Then, in the third and final step, we find the equilibrium emittances from the eigenvalues of $\Sigma S$.

Strictly speaking, since $R$ is not symplectic, the emittances are not invariant
as the bunch moves around the ring.  Therefore, we may expect to find a different
emittance at each point around the ring.  However, if radiation effects are fairly
small, then the variations in the emittances will also be small.

The transfer matrices $R$ and $D$ for an entire ring can be constructed from
the transfer matrices for individual components in the ring.  As an example, we shall consider
a thin `slice' of a dipole.  This is an important case, since
in most storage rings, radiation effects are significant only in dipoles.  Furthermore,
complete dipoles can be constructed by composing the maps for a number of slices.
Hence, once we have a map for a thin slice of dipole, and knowing the usual (symplectic)
transfer maps for drift spaces, quadrupoles and rf cavities, we will be able to construct
the map for one complete turn of a storage ring, starting at any point.

Recall that the transformations for the phase space variables in the
emission of radiation carrying momentum $dP$ are:
\begin{eqnarray}
x & \mapsto & x,
\label{radiationcoordinatetransformation1} \\
p_x & \mapsto & \left( 1 - \frac{dP}{P_0} \right) p_x,  \\
y & \mapsto & y, \\
p_y & \mapsto & \left( 1 - \frac{dP}{P_0} \right) p_y, \\
z& \mapsto & z, \\
\delta & \mapsto & \delta - \frac{dP}{P_0},
\label{radiationcoordinatetransformation6}
\end{eqnarray}
where $P_0$ is the reference momentum.  In general, $dP$ is a function of the co-ordinates.
To find the transformation matrices $R$ and $D$, we find an explicit expression
for $dP/P_0$, and then write down the above transformations to first order.
For an ultra-relativistic particle, the momentum lost through radiation can
be expressed in terms of the synchrotron radiation power $P_\gamma$
(energy loss per unit time):
\begin{equation}
\frac{dP}{P_0} \approx \frac{P_\gamma}{E_0} dt \approx
\frac{P_\gamma}{E_0} \left( 1 + \frac{x}{\rho} \right) \frac{ds}{c},
\end{equation}
where $\rho$ is the radius of curvature of the reference trajectory.
The radiation power $P_\gamma$ is given by (\ref{synchrotronradiationpower}).
In general, the dipole may have a quadrupole gradient, so the field is:
\begin{equation}
B = B_0 + B_1 x.
\end{equation}
Also, the particle may have some energy deviation, so the total energy is:
\begin{equation}
E = E_0 (1 + \delta).
\end{equation}
Substituting these expressions, we find (after some manipulation):
\begin{equation}
P_\gamma = c \frac{C_\gamma}{2\pi}
\left( \frac{1}{\rho^2} + 2k_1 \frac{x}{\rho} \right)
(1 + \delta)^2 E_0^4,
\end{equation}
where $k_1$ is the normalised quadrupole gradient in the dipole:
\begin{equation}
k_1 = \frac{q}{P_0} B_1.
\end{equation}
Hence, the normalised momentum loss may be written:
\begin{equation}
\frac{dP}{P_0} \approx \frac{C_\gamma}{2\pi}
\left( \frac{1}{\rho^2} + 2k_1 \frac{x}{\rho} \right)
\left( 1 + \frac{x}{\rho} \right) (1 + \delta)^2 E_0^3 \, ds.
\end{equation}
Expanding to first order in the phase space variables, we can write:
\begin{equation}
\frac{dP}{P_0} \approx
\frac{C_\gamma}{2\pi} \frac{E_0^3}{\rho^2} \, ds +
\frac{C_\gamma}{2\pi} \left( \frac{1}{\rho^2} + 2k_1 \frac{x}{\rho} \right) \frac{E_0^3}{\rho} x \, ds +
2 \frac{C_\gamma}{2\pi} \frac{E_0^3}{\rho^2} \delta \, ds +
O(x^2) + O(\delta^2).
\label{dpbyp0}
\end{equation}

Given the expression (\ref{dpbyp0}) for $dP/P_0$, the transformations
(\ref{radiationcoordinatetransformation1})--(\ref{radiationcoordinatetransformation6})
become (to first order in the dynamical variables):
\begin{eqnarray}
x & \mapsto & x, \label{radiationcoordinatetransformation1a} \\
p_x & \mapsto & \left( 1 - \frac{C_\gamma}{2\pi} \frac{E_0^3}{\rho^2} \, ds \right) p_x,  \\
y & \mapsto & y, \\
p_y & \mapsto & \left( 1 - \frac{C_\gamma}{2\pi} \frac{E_0^3}{\rho^2} \, ds \right) p_y, \\
z& \mapsto & z, \\
\delta & \mapsto & \left( 1 - 2\frac{C_\gamma}{2\pi} \frac{E_0^3}{\rho^2} \, ds \right)\delta - 
\frac{C_\gamma}{2\pi} \left( \frac{1}{\rho^2} + 2k_1 \frac{x}{\rho} \right) \frac{E_0^3}{\rho} x \, ds -
 \frac{C_\gamma}{2\pi} \frac{E_0^3}{\rho^2} \, ds. \label{radiationcoordinatetransformation6a}
\end{eqnarray}
The first order terms give the components of $R_\textrm{dip}(ds)$, the transfer matrix for
a thin slice (length $ds$) of a dipole.  There is a zeroth order term in the map for the dynamical variables
that will contribute to (the (6,6) component of) $D_\textrm{dip}(ds)$, which contains the zeroth order
terms in the transformation of the Sigma matrix through a thin slice of a dipole.  Since the (6,6) component
of $D_\textrm{dip}(ds)$ represents the quantity $\langle \Delta \delta^2 \rangle$,
the contribution to this component from the zeroth order term in (\ref{radiationcoordinatetransformation6a})
will be second order in $ds$.
We still have to take proper account of the quantum nature of the radiation.  This
will make an additional contribution to $D_\textrm{dip}(ds)$.

The zeroth order term in the map for the Sigma matrix is given by:
\begin{equation}
\left[ D_\textrm{dip}(ds) \right]_{66} = \left\langle \left( \frac{dP}{P_0} \right)^{\!\!2} \right\rangle
\approx \frac{\langle u^2 \rangle}{E_0^2},
\end{equation}
where $\langle u^2 \rangle$ is the mean square of the photon energy.
Using (\ref{integralndotusquared}),
we find that, to zeroth order in the phase space variables:
\begin{equation}
\left\langle \left( \frac{dP}{P_0} \right)^{\!\!2} \right\rangle \approx
2 C_q \gamma^2 \frac{C_\gamma}{2\pi} \frac{E_0^3}{\rho^3} \, ds.
\end{equation}
Note that this term is first order in $ds$, whereas the contribution to $D_\textrm{dip}(ds)$
that we found previously was second order in $ds$.  Hence, in the limit $ds \to 0$,
the latter contribution dominates over the previous contribution.

Collecting the above results, and taking only dominant contributions in the
limit $ds \to 0$, we find that the radiation in a thin slice of dipole has an
effect on the Sigma matrix that can be represented by:
\begin{equation}
\Sigma \mapsto R_\textrm{dip}(ds) \Sigma R_\textrm{dip}\transpose(ds) + D_\textrm{dip}(ds),
\end{equation}
where:
\begin{equation}
R_\textrm{dip}(ds) =
\left(
\begin{array}{cccccc}
1   & 0                                                                              &  0 & 0 &  0 & 0 \\
0   & 1 - \frac{C_\gamma}{2\pi} \frac{E_0^3}{\rho^2} ds   &  0 & 0 &  0 & 0 \\
0   & 0                                                                              &  1 & 0 &  0 & 0 \\
0   & 0                                                                              &  0 & 1 - \frac{C_\gamma}{2\pi} \frac{E_0^3}{\rho^2} ds &  0 & 0 \\
0   & 0                                                                              &  0 & 0 &  1 & 0 \\
- \frac{C_\gamma}{2\pi} \left( \frac{1}{\rho^2} + 2k_1 \right) \frac{E_0^3}{\rho} ds   & 0 &  0 & 0 &  0 & 1 - 2\frac{C_\gamma}{2\pi} \frac{E_0^3}{\rho^2} ds
\end{array} \right),
\end{equation}
and:
\begin{equation}
D_\textrm{dip}(ds) =
\left(
\begin{array}{cccccc}
0 & 0 & 0 & 0 & 0 & 0 \\
0 & 0 & 0 & 0 & 0 & 0 \\
0 & 0 & 0 & 0 & 0 & 0 \\
0 & 0 & 0 & 0 & 0 & 0 \\
0 & 0 & 0 & 0 & 0 & 0 \\
0 & 0 & 0 & 0 & 0 & 2C_q \gamma^2 \frac{C_\gamma}{2\pi} \frac{E_0^3}{\rho^3} ds.
\end{array} \right).
\end{equation}

To construct the full single-turn transformation, we need to compose the maps
for all the elements in the ring, including the radiation effects in the dipoles.
It is straightforward to do this numerically using a computer.  However, some care
is needed in handling the $D$ matrices.
For example, given the Sigma matrix at a location $s_0$, we find the Sigma matrix
at a location $s_1 = s_0 + ds$ from:
\begin{equation}
\Sigma (s_1) = R(s_1; s_0) \Sigma (s_0) R\transpose(s_1; s_0) + D(s_1; s_0).
\end{equation}
Then the Sigma matrix at $s_2$ is given by:
\begin{eqnarray}
\Sigma (s_2) & = & R(s_2; s_1) \Sigma (s_1) R\transpose(s_2; s_1) + D(s_2; s_1) \nonumber \\
& = & R(s_2; s_0) \Sigma (s_0) R\transpose(s_2; s_0) + R(s_2; s_1) D(s_1; s_0) R\transpose(s_2; s_1) + D(s_2; s_1). 
\end{eqnarray}
Hence:
\begin{eqnarray}
R(s_2; s_0) & = & R(s_2; s_1) R(s_1; s_0) \\
D(s_2; s_0) & = & R(s_2; s_1) D(s_1; s_0) R\transpose(s_2; s_1) + D(s_2; s_1).
\end{eqnarray}
Continuing the process, we find:
\begin{eqnarray}
R(s_n; s_0) & = & R(s_n; s_{n-1}) R(s_{n-1}; s_{n-2}) \cdots R(s_1; s_0)\\
            &   & \nonumber \\
D(s_n; s_0) & = & \sum_{r=1}^n R(s_n; s_r) D(s_r; s_{r-1}) R\transpose(s_n; s_r)
\end{eqnarray}
When composing the transfer maps for thin slices of a dipole, we have to remember
to `interleave' the radiation maps with the usual symplectic transport map for a thin slice of dipole.

The next step in finding the equilibrium emittances is to solve the matching condition
(\ref{singleturndampingexcitation}) to find the Sigma matrix for the equilibrium distribution.
To do this (for given
matrices $R$ and $D$), we make use of the eigenvectors $U$ of $R$, and the diagonal
matrix $\Lambda$ constructed from the eigenvalues of $R$:
\begin{equation}
R U = \Lambda U.
\end{equation}
Defining $\tilde{\Sigma}$ and $\tilde{D}$ by:
\begin{eqnarray}
\Sigma & = & U \tilde{\Sigma} U\transpose, \\
D      & = & U \tilde{D} U\transpose,
\end{eqnarray}
the solution for the Sigma matrix can be written:
\begin{equation}
\tilde{\Sigma}_{ij} = \frac{\tilde{D}_{ij}}{1 - \Lambda_i \Lambda_j}.
\end{equation}
The above formulae enable us to find the matched (equilibrium) distribution $\Sigma$;
the eigenvalues of $\Sigma S$ are then $\pm i \varepsilon_k$, where $\varepsilon_k$
are the emittances.

The envelope method makes explicit the fact that vertical emittance is generated by
coupling between the vertical and longitudinal planes in regions where
radiation is emitted (i.e.~by vertical dispersion in dipoles),
and by coupling between the vertical and horizontal planes in regions where
radiation is emitted (i.e.~by betatron coupling in dipoles).
Here, we need to be careful in the use of the term `coupling'.  In this
context, coupling means the presence of non-zero components off the block-diagonals
in the single-turn matrix, $R$.
Full characterisation of the coupling requires complete specification of all
the components off the block-diagonals in $R$.
Depending on these components, it is possible to have coupling in a storage ring,
and not generate any vertical emittance.  For example, one could construct a
closed `coupling bump' using sets of skew quadrupoles in a straight section in a
storage ring.  With proper care in the design, outside the region between the
skew quadrupoles, the vertical motion can be completely decoupled from
the horizontal and the longitudinal.  Then, despite the presence of strong
coupling in some parts of the storage ring, the equilibrium vertical emittance
will come only from the opening angle of the cone describing the spatial
distribution of the synchrotron radiation.

Numerical computational procedures (such as the envelope method) for finding the
equilibrium beam distribution in a storage ring are important because they provide
ways to calculate the equilibrium emittances in complex, coupled lattices.
It is possible to include other non-symplectic effects in the calculation (such as,
for example, intrabeam scattering).

\subsection{Ultra-low emittance tuning}

Often, coupling comes from magnet alignment errors, which will not be known
completely in an operating machine. At the design stage, it is important to
characterise the sensitivity of a lattice to magnet alignment errors, particularly
regarding the vertical emittance.
Being able to compute the beam emittances in a storage ring with coupling
errors present makes it possible to simulate the effects of various types and sizes of
alignment error -- and then to optimise the lattice design to minimise the
sensitivity to the likely errors.  However, in practice, tuning a storage ring to achieve a
vertical emittance of no more than a few
picometres (which may be required for some applications) is a considerable challenge,
even in a lattice designed so as to minimise the sensitivity to coupling errors.
Accurate alignment (by survey) of the magnets is always the first step in achieving
ultra-low emittances; but beam-based tuning methods will then also be needed.

\begin{figure}[t]
\begin{center}
\includegraphics[width=0.55\textwidth]{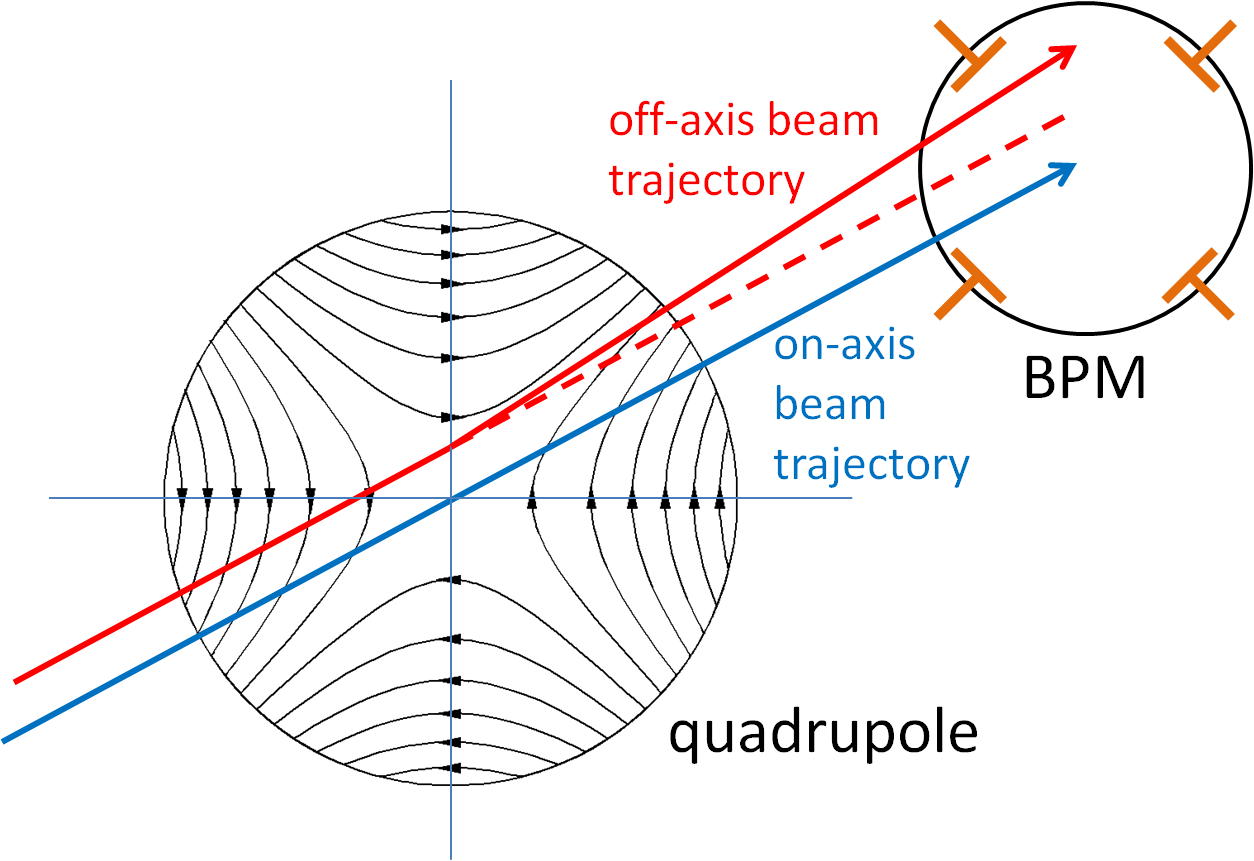}
\caption{Beam-based alignment in a quadrupole.  If the beam passes
off-axis through a quadrupole magnet, then varying the strength of
the magnet changes the trajectory downstream of the magnet.  A
change in trajectory can be observed in a beam position monitor (BPM).
One method of beam-based alignment consists of steering the beam
(using upstream orbit corrector magnets) until changing the quadrupole
strength has no effect on the beam position observed in the BPM.}
\label{figbeambasedalignment}
\end{center}
\end{figure}

A variety of beam-based methods for tuning storage rings have been developed
over the years.
A typical procedure might look as follows:
\begin{enumerate}
\item Align the magnets by a survey of the ring.
Typically, quadrupoles need to be aligned to better than a few tens of microns, and
sextupoles to better than a couple of hundred microns.
\item Determine the positions of the BPMs relative to the quadrupoles.
This is known as `beam-based alignment' (BBA, see Fig.~\ref{figbeambasedalignment}),
and can be achieved by steering the beam to a position in each quadrupole
where changing the quadrupole strength has no effect on the orbit \cite{zimmermannminty2003}.
\item Correct the orbit (using steering magnets) so that it is as close as
possible to the centres of the quadrupoles.
\item Correct the vertical dispersion (using steering magnets and/or skew
quadrupoles, and measuring the dispersion at the BPMs) as close to zero as possible.
\item Correct the coupling, by adjusting skew quadrupoles so that an orbit `kick'
in one plane (from any orbit corrector) has no effect on the orbit in the other plane.
\end{enumerate}
Usually, the last three steps need to be iterated several (or even many) times.

Results from the tuning procedure described above can be limited by systematic
errors on the BPMs, which can affect dispersion and coupling measurements.
A useful technique for overcoming such limitations is to apply Orbit Response
Matrix (ORM) analysis \cite{safranek1997}.  This can be used to determine a wide
range of magnet and diagnostics parameters, including coupling errors and BPM tilts.
Although vertical emittances of order 1\,pm have now been achieved (representing
an emittance ratio of less than 0.1\%), tuning an electron storage ring to operate
in this regime still remains a challenging goal, requiring extensive work and application
of a range of techniques to reduce errors.  Even making measurements of emittances
less than a few picometres is not straightfoward, and requires specialist instrumentation
(see, for example, \cite{andersson2008}).

\end{document}